\documentclass[12pt,a4paper]{article}
\pdfoutput=1
\usepackage{graphicx}
\usepackage[T1]{fontenc}
\usepackage[utf8]{inputenc}
\usepackage{textcomp}
\usepackage[sc,osf]{mathpazo}
\usepackage{a4wide}  
\usepackage{latexsym,amsthm,amsfonts,amsmath,mathrsfs,amssymb}
\usepackage{dsfont}
\usepackage{accents}
\usepackage[nosort]{cite}
\usepackage{booktabs} 
\usepackage[unicode,implicit]{hyperref}
\hypersetup{%
  pdftitle    = {alpha' corrections of Reissner-Nordstrom black holes}
  pdfkeywords = {black hole, string theory, entropy, temperature, Wald, Hawking,
    branes, supergravity, supersymmetry, corrections, WGC, weak gravity
    conjecture, heterotic superstring},
  pdfauthor   = {Pablo A. Cano, Samuele Chimento, Roman Linares, Tomas Ortin,
    Pedro F. Ramirez},
  plainpages  = true,
  colorlinks  = true,
  citecolor   = blue,
  urlcolor    = red,
  linkcolor   = black
}
\newcommand{\hepth}[1]{{\tt
\href{http://www.arXiv.org/abs/hep-th/#1}{hep-th/#1}}}
\newcommand{\grqc}[1]{{\tt
\href{http://www.arXiv.org/abs/gr-qc/#1}{gr-qc/#1}}}

\newcommand{\arxiv}[1]{{\tt arXiv:\href{http://www.arXiv.org/abs/#1}{#1}}}
\newcommand{\doi}[1]{{\tt DOI:\href{http://dx.doi.org/#1}{#1}}}

\makeatletter
\@addtoreset{equation}{section}
\makeatother

\begin{document}

\vspace{-0.5cm}
\begin{flushright}
\small
IFT-UAM/CSIC-18-114\\
October 22\textsuperscript{nd}, 2019\\
\normalsize
\end{flushright}

\vspace{.2cm}

\begin{center}

{\Large {\bf {$\alpha'$ corrections of Reissner-Nordstr\"om black holes}}}

\vspace{0.8cm}

\renewcommand{\thefootnote}{\alph{footnote}}
{\sl\large Pablo A.~Cano,}$^{1,2,}$\footnote{\tt pabloantonio.cano[at]kuleuven.be}
{\sl\large Samuele Chimento,}$^{1,}$\footnote{\tt  samuele.chimento[at]csic.es}
{\sl\large Rom\'an Linares,}$^{3,}$\footnote{\tt lirr[at]xanum.uam.mx}\\[.5cm]
{\sl\large Tom\'{a}s Ort\'{\i}n}$^{1,}$\footnote{\tt tomas.ortin[at]csic.es}
{\sl\large  and  Pedro F.~Ram\'{\i}rez}$^{4,}$\footnote{\tt pedro.ramirez[at]aei.mpg.de}

\setcounter{footnote}{0}
\renewcommand{\thefootnote}{\arabic{footnote}}

\vspace{0.5cm}

${}^{1}${\it Instituto de F\'{\i}sica Te\'orica UAM/CSIC\\
C/ Nicol\'as Cabrera, 13--15,  C.U.~Cantoblanco, E-28049 Madrid, Spain}\\ 

\vspace{0.5cm}
${}^{2}${\it Instituut voor Theoretische Fysica, KU Leuven\\
	Celestijnenlaan 200D, B-3001 Leuven, Belgium}\\

\vspace{0.5cm}

${}^{3}${\it Departamento de F\'{\i}sica, Universidad Aut\'onoma Metropolitana\\ 
Iztapalapa, San Rafael Atlixco 186, C.P. 09340 Mexico City, Mexico}

\vspace{0.5cm}

${}^{4}${\it Max-Planck-Institut f\"ur Gravitationsphysik (Albert-Einstein-Institut), \\
Am M\"uhlenberg 1, D-14476 Potsdam, Germany}

\vspace{0.8cm}


{\bf Abstract}
\end{center}
\begin{quotation}
  {\small We study the first-order in $\alpha'$ corrections to non-extremal
    4-dimensional dyonic Reissner-Nordstr\"om (RN) black holes with equal
    electric and magnetic charges in the context of Heterotic Superstring
    effective field theory (HST) compactified on a $T^{6}$. The particular
    embedding of the dyonic RN black hole in HST considered here is not
    supersymmetric in the extremal limit. We show that, at first order in $\alpha'$,
    consistency with the equations of motion of the HST demands additional
    scalar and vector fields become active, and we provide explicit expressions for all of them. We determine analytically the
    position of the event horizon of the black hole, as well as the corrections to the extremality bound,
    to the temperature and to the entropy, checking that they are related by
    the first law of black-hole thermodynamics, so that
    $\partial S/\partial M=1/T$. We discuss the implications of our results in
    the context of the Weak Gravity Conjecture, clarifying that entropy corrections for fixed mass and charge at extremality do not necessarily imply corrections to the extremal charge-to-mass ratio.
  }
\end{quotation}

\newpage
\pagestyle{plain}

\tableofcontents


\section{Introduction}

Superstring Theory (ST) is our prime candidate for a consistent theory of
quantum gravity. One of the main applications of such a theory would be the
study of black holes and their quantum behaviour. Thus, it is hardly
surprising that one of the main areas of research in ST is black-hole physics:
construction of black-hole solutions, calculation of their Hawking temperature
and Bekenstein-Hawking entropy, the microscopic interpretation of the latter,
etc. 


A large part of all this research relies on the effective field theory actions
that describe the low-energy behavior of ST and which are (if the vacua chosen
preserve any supersymmetries) standard supergravity theories plus terms of
higher order in the Regge slope parameter $\alpha'$ (and, correspondingly, in
curvatures) and in the string coupling constant.\footnote{The addition of
  these terms still leaves us with (less standard) supergravity theories, with
  terms of higher order in curvatures, but supergravity theories nevertheless,
  since supersymmetry should be preserved. Often, the $\alpha'=0$ limit is
  improperly referred to as the ``supergravity limit'', though.} The terms of
higher order in $\alpha'$ are important from the ST point of view because they
represent genuine stringy departures of matter-coupled General Relativity due
to the non-vanishing string`s length $\ell_{s}$, with $\alpha'=\ell_{s}^{2}$.

Most of the stringy black-hole solutions constructed in the literature,
though, only solve the zeroth-order limit of these effective field
theories. This means that they can only be considered good ST solutions if one
can prove that taking into account the terms of higher order in $\alpha'$ only
introduces small corrections in the solutions. Because of the technical
complications involved in dealing with higher-order actions, only an
estimation of the size of these corrections based on the values of curvature
scalars evaluated over the zeroth-order solution are typically made. Quite frequently, it is possible to minimize these scalars by constraining the relative values of the black hole parameters, hoping that any possible stringy effects are also minimized.

However, we are learning that the introduction of higher-curvature terms can have important physical consequences which one cannot make disappear by simply decreasing the curvature scalars. In a perturbative approach, these corrections can be interpreted as introducing delocalized sources in the equations of motion, which may contribute to the global charges and energy of the system \cite{Sen:1997zb, Cano:2018qev,Cano:2018brq, Faedo:2019xii}. Moreover, as it was shown in Refs.~\cite{Cano:2018qev,Cano:2018brq} using the results
of Ref.~\cite{Chimento:2018kop} and by direct computation of the
$\alpha'$ corrections of some supersymmetric black-hole solutions, the
curvature scalars do not capture all the possible non-vanishing terms than
can occur in the equations of motion at higher orders in $\alpha'$. While the higher-curvature terms are relevant in any configuration, it has been shown that, in some special situations, taking them into account is just fundamental --- see \cite{Cano:2018aod,Cano:2018hut}. The inevitable conclusion is that very relevant information can be acquired by performing explicit calculations of the $\alpha'$ corrections to the zeroth-order black-hole solutions. Our goal in this paper is to extend the results found in Refs.~\cite{Cano:2018qev,Cano:2018brq, Cano:2019oma} to
non-supersymmetric and non-extremal black holes and compute explicitly their
first-order $\alpha'$ corrections in some consistent ST effective action
framework.

Corrections to non-extremal, uncharged, rotating black holes (non-extremal
Kerr black holes) have been studied long ago, in
Refs.~\cite{Campbell:1990ai,Campbell:1990fu,Mignemi:1992pm}, where it was
shown that, at first order in $\alpha'$, stringy fields different from the
metric are activated (the dilaton and the Kalb-Ramond 2-form).\footnote{On the other hand, the backreaction of those fields onto the metric appear at order $\alpha'^2$, and such corrections have been more recently computed in the stringy-inspired Einstein-dilaton-Gauss-Bonnet and dynamical Chern-Simons theories -- see \text{e.g.}  \cite{Yunes:2009hc,Yunes:2011we,Pani:2011gy,Ayzenberg:2014aka,Cano:2019ore}}
In this work we address
this problem for 4-dimensional, charged, non-rotating, non-extremal,
(Reissner-Nordstr\"om (RN)) black holes embedded in ST.

One of the lessons learned in Refs.~\cite{Cano:2018qev,Cano:2018brq} is that
$\alpha'$ corrections to 4- and 5-dimensional systems can be conveniently computed directly in $d=10$, without having to make
any assumptions or approximations, especially in the
framework of the Heterotic Superstring Theory (HST) effective field
theory. One just needs to find the 10-dimensional solution whose dimensional
reduction gives rise to the black hole (or other) solution under
consideration, if it exists. Otherwise, the lower-dimensional solution is not
a ST theory solution and computing its $\alpha'$ corrections is meaningless.

Since RN black holes are not purely gravitational (there is, at least, one
vector field active), there is more than one embedding of the 4-dimensional RN
black hole in 10-dimensional HST, corresponding to the many ways in which the
vector field can be obtained from the 10-dimensional fields: as Kaluza-Klein
or winding vector fields, from 10-dimensional vector fields etc. An important
difference between the possible embeddings is the the amount of unbroken
supersymmetries of their extremal limits. For instance, the embedding of the
extremal RN black hole considered in Ref.~\cite{Cano:2018brq} preserved half
of the possible supersymmetries unbroken. Here we are going to consider an
embedding which breaks all supersymmetries in the extremal limit, where it
will coincide, up to T-duality transformations, with the embedding found in
Ref.~\cite{Khuri:1995xq}. This embedding is described in
Section~\ref{sec-nonsusyRNBH}.  In Section~\ref{sec-corrections} we will
describe the first-order $\alpha'$ corrections for the 10-dimensional solution
that gives rise to this 4-dimensional RN black hole and we will dimensionally
reduce the 10-dimensional configuration to recover the 4-dimensional
$\alpha'$-corrected fields (the calculations are described in the
appendices). Then, we determine the position of the event horizon of the
corrected solution in Section~\ref{sec-horizons}, its temperature in
Section~\ref{sec-temperature} and its Wald entropy in
Section~\ref{sec-entropy}. We discuss our results and describe their relation with the WGC in
Section~\ref{sec-discussion}.

\section{A non-supersymmetric dyonic Reissner-Nordstr\"om black hole}
\label{sec-nonsusyRNBH}

Our starting point is a zeroth-order in $\alpha'$ solution of the
10-dimensional Heterotic Superstring effective field theory
(HST)\footnote{the action and equations of motion of this theory are
  described in Appendix~\ref{sec-heteroticalpha}.} given by the
following 10-dimensional fields, which we distinguish from the
4-dimensional ones by the hats:\footnote{Using the components of the
  Ricci tensor etc.~computed in Appendix~\ref{sec-connections}, it
  takes little time to check that it satisfies
  Eqs.~(\ref{eq:eq1})-(\ref{eq:eq3}) at zeroth order in $\alpha'$.}:

\begin{equation}
\label{eq:RNBH}
\begin{array}{rcl}
d\hat{s}^{2}
& = &
{\displaystyle
a^{2} dt^{2} - \frac{dr^{2}}{a^{2}} 
-r^{2}[d\theta^{2} +\sin^{2}{\theta}d\phi^{2}]
-dz^{2} -d\vec{y}^{\, 2}_{(5)}\, ,
}
\\
& & \\
\hat{\phi}
& = & 
\hat{\phi}_{\infty}\, ,
\\    
& & \\
\hat{H}
& = &
e (dt\wedge dr + r^{2}\sin{\theta} d\theta \wedge d\phi) \wedge dz\, .
\end{array}
\end{equation}

\noindent
The functions $a(r)$ and $e(r)$ are given by

\begin{equation}
\label{eq:aefunctions}
a^{2} = 1 -\frac{2M}{r} +\frac{p^{2}/2}{r^{2}}\, ,
\hspace{1.5cm}
e = \frac{p}{r^{2}}\, ,  
\end{equation}

\noindent
and $\phi_{\infty}, p, M$ are physical constants. 

The above 10-dimensional metric is the direct product of that of the
4-dimensional, non-extremal RN black hole of mass $M$ and that of a
flat $T^{6}$. A trivial dimensional reduction on that $T^{6}$ (with
coordinates $z, y^{1},\cdots,y^{5}$), gives the 4-dimensional metric
of that black hole and no additional, active, Kaluza-Klein vector or
scalar fields.

The function $a$ that characterized the RN black hole metric can be
rewritten in the form

\begin{equation}
a^{2} = \frac{(r-r_{+})(r-r_{-})}{r^{2}}\, ,  
\end{equation}

\noindent
where, as usual,

\begin{equation}
  \label{eq:rpm}
r_{\pm}= M\pm\sqrt{M^{2}-p^{2}/2}\, ,
\end{equation}

\noindent
are the values of $r$ at which the outer ($+$) and inner ($-$)
horizons are placed, if they are real, as we are going to assume here.

On the other hand, the dimensional reduction of the Kalb-Ramond 2-form
only gives a dyonic vector field $B_{\mu}$ with equal (up to signs)
electric and magnetic charges (proportional to the constant $p$ in the
solution), whose field strength $F(B)_{\mu\nu}$ squares to
zero.\footnote{This property follows trivially from $\hat{H}^{2}=0$.}
It is this property that allows us to have a constant Kaluza-Klein
scalar in the $z$ direction, since the equation of motion of that
scalar would be $\nabla^{2}k\sim F^{2}$. The dilaton field is also
constant in $d=4$.

As we have mentioned in the introduction, in the extremal limit
$M=|p|/\sqrt{2}$ the 10-dimensional solution is T-dual in the $z$
direction to the non-supersymmetric, purely gravitational solution
found in Ref.~\cite{Khuri:1995xq} and, therefore, it is not
supersymmetric. Being related by T-duality, these two 10-dimensional
solutions give rise to the same 4-dimensional RN black hole.

\section{$\alpha'$ corrections}
\label{sec-corrections}

In order to find the $\alpha'$ corrections to this solution, we have
to use an ansatz that can accommodate both the above solution and the
potential $\alpha'$ corrections, which may activate other components
of the 10-dimensional metric or the Kalb-Ramond field or the dilaton
\cite{Campbell:1990ai,Campbell:1990fu,Mignemi:1992pm}. If the ansatz
is not general enough, it will not be possible to solve all the
equations of motion and it will be necessary to add to it further
active components to be determined.

After several trials, we have arrived, for the zeroth-order solution
that we are considering, to the following ansatz:

\begin{equation}
  \label{eq:ansatzcorrections1}
\begin{array}{rcl}
d\hat{s}^{2} 
& =  &
A^{2}dt^{2} - B^{2}dr^{2} - r^{2}[d\theta^{2}+\sin^{2}{\theta}d\phi^{2}]
-C^{2}[dz+Fdt]^{2} - d\vec{y}^{\, 2}_{5}\, ,
\\
& & \\
\hat{\phi}
& = &
\hat{\phi}_{\infty} +\alpha'\delta_{\phi}\, ,
\\
& & \\
\hat{H}
& = & 
D\, \hat{e}^{\, 0} \wedge \hat{e}^{\, 1}\wedge \hat{e}^{\, 4} 
+E\, \hat{e}^{\, 2} \wedge \hat{e}^{\, 3} \wedge \hat{e}^{\, 4}
+G\, \hat{e}^{\, 0} \wedge \hat{e}^{\, 2}\wedge \hat{e}^{\, 3}\, ,
\end{array}
\end{equation}

\noindent
where the Zehnbein 1-forms $\hat{e}^{\, a}$ are defined in
Eq.~(\ref{eq:Vielbein}) and where $A,B,C,D,E,F,G$ and $\delta_{\phi}$ are
functions of the coordinate $r$. The expansion of the 7 functions
$A,B,C,D,E,F,G$ in powers of $\alpha'$ is assumed to be of the form

\begin{equation}
  \label{eq:ansatzcorrections2}
  \begin{array}{rcl}
A & \sim &a+\alpha' \delta_{A}\, ,\,\,\,\,\,\,\,
B \sim a^{-1}+\alpha' \delta_{B}\, ,\,\,\,\,\,\,
C\sim 1 +\alpha' \delta_{C}\, ,\,\,\,\,\,\,\,
F\sim \alpha' \delta_{F}\, ,  
\\
& & \\
D & \sim & e +\alpha'\delta_{D}\, ,\,\,\,\,\,\,
E\sim e +\alpha'\delta_{E}\, ,\,\,\,\,\,\,
G\sim\alpha'\delta_{G}\, ,  
\end{array}
\end{equation}

\noindent
where the functions $a$ and $e$ are those present in the zeroth-order solution
Eq.~(\ref{eq:aefunctions}).

Thus, setting $\alpha'=0$ in the above configuration
Eqs.~(\ref{eq:ansatzcorrections1}) we recover the RN solution
Eq.~(\ref{eq:RNBH}) and the 8 functions $\delta_{A}$, $\delta_{B}$,
$\delta_{C}$, $\delta_{D}$, $\delta_{E}$, $\delta_{F}$, $\delta_{G}$ and
$\delta_{\phi}$ describe the first-order $\alpha'$ corrections to that
solution.

The details of the procedure we have followed to find these
corrections can be found in 
Appendix~\ref{sec-solutioncorrections}. Here we are just going to
quote the results in 4-dimensional language (unhatted fields),
stressing that we have determined the new integration constants by
demanding that the mass is not renormalized by the $\alpha'$
corrections\footnote{This is equivalent to considering the $M$ that
  appears in the corrected solutions as the renormalized mass.} and
that the fields are regular at the outer (event) horizon at $r_{+}$
since it is not possible to keep them regular at both $r_{+}$ and
$r_{-}$ (which are assumed to be different in this calculation)
simultaneously. The singularity of the scalar fields at $r_{-}$ is
clearly related to the instability of the Cauchy horizon.

First of all, observe that, once the dilaton and the Kaluza-Klein scalar
measuring the size of the $S^{1}$ parametrized by the coordinate $z$,
$k\equiv |\hat{g}_{zz}|^{1/2}$, are activated, the 4-dimensional metric in the
Einstein frame will be given by

\begin{eqnarray}
ds^{2}
& = &
Ce^{-2(\hat{\phi}-\hat{\phi}_{\infty})}\left[A^{2}dt^{2} - B^{2}dr^{2}
      -r^{2}(d\theta^{2}+\sin^{2}{\theta}d\varphi^{2})\right]
      \nonumber \\
  & & \nonumber \\
  & \equiv &
             \label{eq:correcterdmetric1}
           N^{2} f dt^{2}-\frac{d\rho^{2}}{f}
           -\rho^{2}(d\theta^{2}+\sin^{2}{\theta}d\varphi^{2})\, ,
\end{eqnarray}

\noindent
where we have defined the new coordinate $\rho$ by

\begin{equation}
\rho=r C^{1/2}e^{-(\hat{\phi}-\hat{\phi}_{\infty})}\, ,
\end{equation}

\noindent
and two new functions $N$ and $f$ in terms of which the equations (and the
solutions) take a surprisingly far simpler form:

\begin{eqnarray}
\label{eq:correcterdmetric2}
N^{2}
  & = &
        1+\alpha' \frac{p^{2}/8}{\rho^{4}}\, ,
  \\
  & & \nonumber \\
  \label{eq:correcterdmetric3}
f
  & = &
        1-\frac{2M}{\rho}+\frac{p^{2}/2}{\rho^{2}}
        -\alpha' \frac{p^{2}/4}{\rho^{4}}\left[1 -\frac{3M/2}{\rho}
        +\frac{11p^{2}/40}{\rho^{2}}\right]\, . 
\end{eqnarray}

\begin{figure}[t!]
\centering
\includegraphics[width=0.7\textwidth]{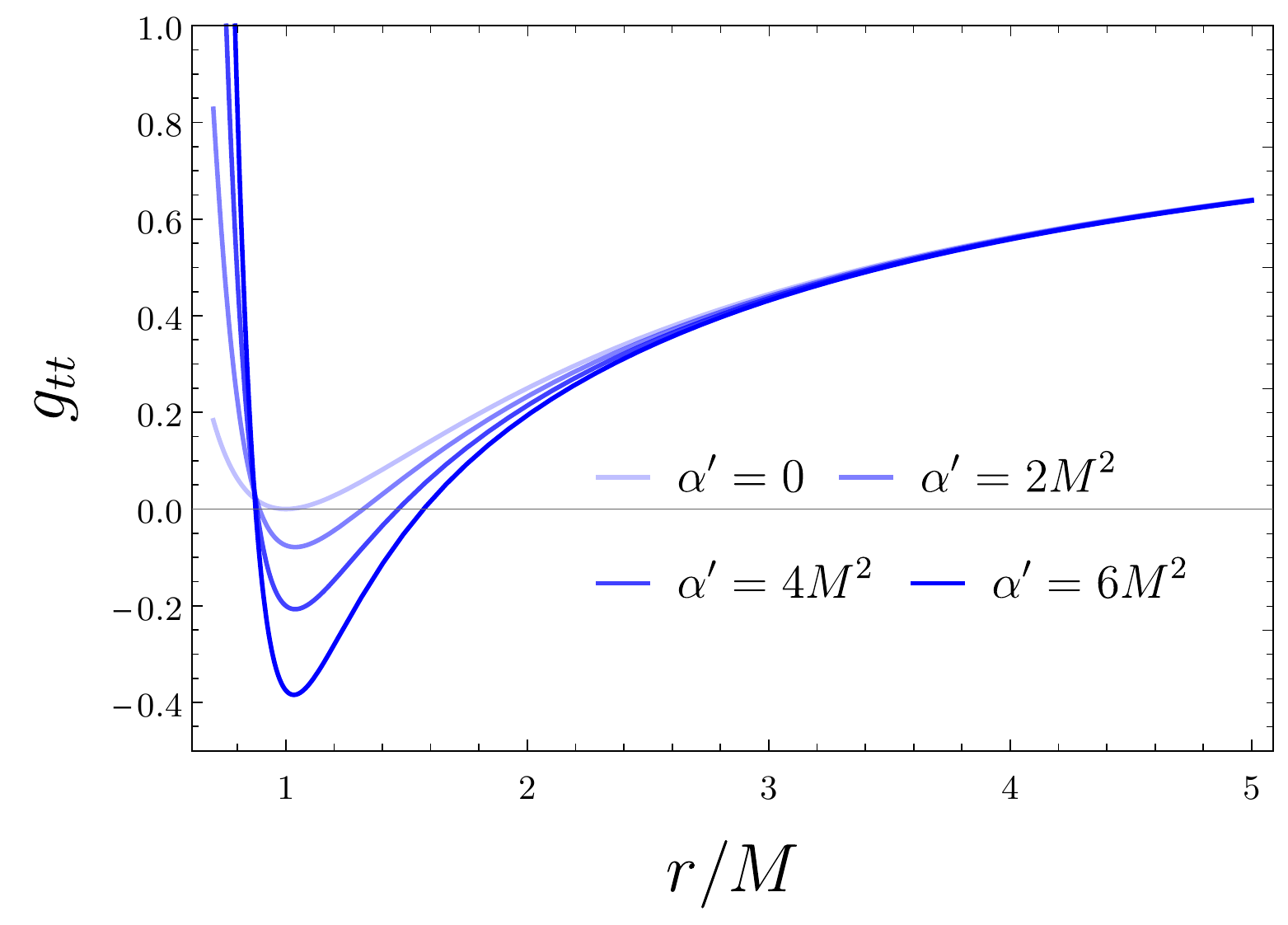}
\caption{\small Profile of $g_{tt}(r)$ for the $\alpha'$-corrected Reissner-Norstr\"om black hole corresponding to the case $M=p/\sqrt{2}$. $\alpha'$-corrections take the black hole away from extremality and increase the size of the outer horizon. The values of $\alpha'/M^2$ chosen are somewhat exaggerated for illustration purposes. }
\label{fig:gtt}
\end{figure}

Observe that the two radial coordinates coincide at zeroth order:

\begin{equation}
r= \rho +\mathcal{O}(\alpha')\, ,  
\end{equation}

\noindent
and that the $\alpha'$ corrections vanish asymptotically, quite
fast. As we will discuss later on, the corrections
start becoming dominant for very small values of the radial
coordinate, typically well inside the inner horizon. In Fig.~\ref{fig:gtt} we show
the profile of $g_{tt}=N^2f$ for the solution with $M=p/\sqrt{2}$, corresponding to
the extremal case at zeroth order. We observe that $\alpha'$-corrections take 
this solution away from extremality, a fact that we will study in more detail in the
next sections. 

The rest of the 4-dimensional fields which are active include, apart
from the the dilaton $e^{-\phi}$ and the Kaluza Klein scalar $k$ that
we have mentioned above, a Kaluza-Klein vector field $A_{\mu}$, a
winding\footnote{This is a vector field that is part of the
  10-dimensional Kalb-Ramond 2-form, while the Kaluza-Klein vectors
  are part of the 10-dimensional metric.} vector field $B_{\mu}$ and a
Kalb-Ramond 2-form $B_{\mu\nu}$ that, in 4-dimensions, can be traded
by an axion field that we are going to denote by $\chi$. They take a
much more complicated forms than the metric, with logarithmic
divergences at $r=r_{-}$. In order to describe them, we first write
them in terms of a minimal number of functions and corrections whose
value can be found in Appendix~\ref{sec-solutioncorrections}.

First of all, using the relation Eq.~(\ref{eq:deltaDdeltaphi}),
$2\delta_{\phi}=r^{2}\delta_{D}/p$, the 4-dimensional Kaluza-Klein scalar and
dilaton fields are given by

\begin{eqnarray}
 k
  & = &
        1+\alpha' \delta_{C}\, ,
        \\
  & & \nonumber \\
        e^{-\phi}
  & = &
        e^{-\hat{\phi}_{\infty}}
        \left[1-\frac{\alpha'}{2p}(r^{2}\delta_{D}- p\delta_{C}) \right]\, .
\end{eqnarray}

The field strengths of the Kaluza-Klein vector field ($A$) and of the
vector field that originates in the 10-dimensional Kalb-Ramond 2-form
($B$) are given by

\begin{eqnarray}
  F(A)
  & = &
      -\alpha' \delta_{F}' dt\wedge d\rho
  \\
  & & \nonumber \\
  F(B)
  & = &
        e 
        \left[1+\alpha'(\delta_{N}+\delta_{D}/e)\right]
        dt\wedge  d\rho
        \nonumber \\
  & & \nonumber \\
  & &  
        +e\rho^{2}\left[1+\alpha'\delta_{E}/e\right]
        \sin{\theta}d\theta \wedge d\phi\, ,
\end{eqnarray}

Finally, the 4-dimensional 3-form field strength is given by

\begin{equation}
H = \alpha' a \delta_{G} r^{2}\sin{\theta} dt\wedge d\theta \wedge d\phi\, .
\end{equation}

As we have already mentioned, in 4 dimensions, the Kalb-Ramond 2-form
can be traded by the axion field $\chi$ that, in this case, would only
depend on the coordinate $r$:

\begin{equation}
d\chi = -\alpha' \delta_{G}d\rho\, .  
\end{equation}

The expressions for all these fields are quite involved and exhibit
logarithmic divergencies at $r=r_{-}$, but we can compute their
charges, defined asymptotically ($\rho\rightarrow \infty$) by

\begin{eqnarray}
  F(A,B)
  & \sim &
           \frac{Q_{A,B}}{\rho^{2}}dt\wedge d\rho
           +P_{A,B}\sin{\theta} d\theta\wedge d\phi\, ,
  \\
  & & \nonumber \\
  e^{\phi}
  & \sim &
           e^{\phi_{\infty}}\left[1 +\frac{Q_{\phi}}{\rho}\right]\, ,
  \\
  & & \nonumber \\
  k
  & \sim &
  1+\frac{Q_{k}}{\rho}\, ,
  \\
  & & \nonumber \\
  d\chi
  & \sim &
  -\frac{Q_{\chi}}{\rho^{2}}d\rho\, .         
\end{eqnarray}

We readily get

\begin{eqnarray}
  Q_{k}
  & = &
        \frac{\alpha' r_{-} \left(140 r_{+}^3-154 r_{+}^{2} r_{-}+35 r_{+} r_{-}^{2}-9 r_{-}^3\right)}{140 r_{+}^3 \left(r_{+}^{2}+4 r_{+} r_{-}+r_{-}^{2}\right)}
        +\mathcal{O}(\alpha^{\prime\, 2})\, ,
  \\
  & & \nonumber \\
  Q_{\phi}
  & = &
        \alpha'\frac{16r_{-}^{4}-77 r_{-}^{3}r_{+}
        -49r_{-}^{2}r_{+}^{2} +70r_{+}^{3}(r_{-}+r_{+})}{280 r_{+}^{3}(r_{-}^{2}+4r_{-}r_{+}+r_{+}^{2})}\, ,
  \\
  & & \nonumber \\
  Q_{\chi}
  & = &
        0+\mathcal{O}(\alpha^{\prime\, 2})\, ,
  \\
  & & \nonumber \\
  Q_{A}
  & = &
        \frac{\alpha' r_{-}(r_{+}-r_{-})^{2}}{4\sqrt{2r_{+}r_{-}}r_{+}^{3}}
        +\mathcal{O}(\alpha^{\prime\, 2})\, ,
  \\
  & & \nonumber \\
  P_{A}
  & = &
        0\, ,
  \\
  & & \nonumber \\
  Q_{B}
  & = &
        p\, ,       \\
  & & \nonumber \\
  P_{B}
  & = &
        p\, .
\end{eqnarray}

A few comments are in order. First, we note that there are no new
independent charges, as all of them are completely determined by $M$
and $p$.  In
the uncharged limit $p\rightarrow 0$ we see that the metric reduces to
the Schwarzschild one, and the only nonvanishing charge is $Q_{\phi}$:

\begin{eqnarray}
  Q_{k}
  & = &
        \frac{\alpha'}{2M}\left\{(p/M)^{2}+\mathcal{O}\left((p/M)^{3}\right)\right\}\, ,
  \\
  \nonumber \\
  Q_{\phi}
  & = &
        \frac{\alpha'}{8M}\left\{1 +\tfrac{1}{4}(p/M)^{2}
        +\mathcal{O}\left((p/M)^{4}\right)\right\}\, ,
  \\
  \nonumber \\
  Q_{A} & = &
              \frac{\alpha'}{32M} \left\{(p/M)
              +\mathcal{O}\left((p/M)^{2}\right)\right\} \, ,
\end{eqnarray}

This is in agreement with previous computations of corrections in
uncharged solutions
\cite{Campbell:1990ai,Campbell:1990fu,Mignemi:1992pm,Yunes:2011we,Pani:2011gy}.

\section{Horizons}
\label{sec-horizons}

Let us now study the horizons of the $\alpha'$-corrected metric
determined by
Eqs.~(\ref{eq:correcterdmetric1}),(\ref{eq:correcterdmetric2}) and
(\ref{eq:correcterdmetric3}).  The horizons of the metric are
determined by the zeroes of the function $f(\rho)$, which, using the
definitions

\begin{equation}
  x\equiv \rho/M\, ,
  \hspace{1cm}
  q\equiv p/(\sqrt{2}M)\, ,
  \hspace{1cm}
  \alpha \equiv \alpha^{\prime}/M^{2}\, ,
\end{equation}

\noindent
can be written in the form

\begin{equation}
  f(x)=  1-\frac{2}{x}+\frac{q^{2}}{x^{2}}
  -\frac{\alpha q^{2}}{2x^{4}}
  +\frac{3\alpha q^{2}}{4x^{5}}  -\frac{11\alpha q^{4}}{40x^{6}}\, .
\end{equation}

As expected, the $\alpha'$ corrections become dominant for small
values of $x$, well inside the inner (Cauchy) horizon. In the
uncorrected RN black hole, $f(x)$ is positive inside this horizon and
diverges when $x$ approaches the singularity at $x=0$. In the
$\alpha'$-corrected RN black hole, when we move towards $x=0$ from the
Cauchy horizon, $f(x)$ is positive, but it always reaches a maximum
and starts decreasing so that $\lim_{x\rightarrow
  0+}f(x)=-\infty$. Therefore, a generic feature of the corrected
black holes is that they have a third horizon inside the Cauchy
horizon,\footnote{Or a second horizon inside the event horizon in the
  extremal case in which the two outermost horizons coincide.}
although it is always placed close to the region at which the
curvature becomes so large that higher corrections in $\alpha'$ can no
longer be ignored.

In order to find the corrections to the positions of the horizons in
the $\alpha'$-corrected RN black holes, we can study the zeroes of the
6\textsuperscript{th}-order polynomial $P(x)\equiv x^{6}f(x)$

\begin{equation}
  \label{eq:P(x)}
P(x)=  x^{6}-2x^{5}+q^{2}x^{4}-\tfrac{1}{2}\alpha q^{2} x^{2}
  +\tfrac{3}{4} \alpha q^{2}x  -\tfrac{11}{40}\alpha q^{4}\, ,
\end{equation}

\noindent
to first order in $\alpha$.

Let us start with the extremal case in which

\begin{equation}
q = 1+a\alpha\, ,  
\end{equation}

\noindent
for some numerical constant $a$. Notice that the charge-to-mass ratio $q$ can differ from 1 in the extremal limit once higher-curvature interactions are incorporated \cite{Campanelli:1994sj,Kats:2006xp, Loges:2019jzs, Cano:2019oma}, which has been connected to the weak gravity conjecture \cite{ArkaniHamed:2006dz}. At first order in $\alpha$ we have

\begin{equation}
  \label{eq:P(x)extremal}
  P(x)
  =
  x^{6}-2x^{5}+(1+2a\alpha)x^{4}-\tfrac{1}{2}\alpha  x^{2}
  +\tfrac{3}{4} \alpha x  -\tfrac{11}{40}\alpha
  +\mathcal{O}(\alpha^{\prime\, 2})\, .
\end{equation}

The numerical results obtained for several values of $q$ suggest that,
in the non-superextremal cases, there is a complex pole and its
conjugate plus a pole at a negative value of $x$, so that, in the
extremal limit, it should be possible to factorize $P(x)$ as follows

\begin{equation}
  P(x)
  =
  |x -(b\alpha+ic\alpha^{1/4})|^{2}
  (x+d\alpha^{1/4}) (x-e\alpha^{1/4})
  [x-(1+f\alpha) ]^{2}
  +\mathcal{O}(\alpha^{2})\, ,
\end{equation}

\noindent
for constants $b,c,d,e,f,g$ to be determined by comparison with
Eq.~(\ref{eq:P(x)extremal}). We readily find the values of the two
constants which determine the corrections to the extremality relation
between the charge and mass and to the position of the horizon

\begin{equation}
  a = 1/80\, ,
  \hspace{.5cm}
  f = 3/40\, ,
\end{equation}

\noindent
so that, 

\begin{eqnarray}
  \label{eq:Mext}
  M_{\rm ext}
  & = &
        (p/\sqrt{2})
        \left[1 -\frac{\alpha'}{80}\frac{1}{(p/\sqrt{2})^{2}} \right]
        +\mathcal{O}(\alpha^{\prime\, 2})\, ,
  \\
  & & \nonumber \\
  \rho_{\rm h\, ext}
  & = &
        M_{\rm ext} \left(1+\frac{3\alpha'}{40M_{\rm ext}^{2}}\right)
       + \mathcal{O}(\alpha^{\prime\, 2})
        =
        (p/\sqrt{2})
        \left[1 +\frac{\alpha'}{16}\frac{1}{(p/\sqrt{2})^{2}} \right]
        +\mathcal{O}(\alpha^{\prime\, 2})\, .
\end{eqnarray}

\noindent
This expression for the correction to the mass in the extremal limit was anticipated in \cite{Cano:2019oma}.

Far from extremality, it should be possible to factorize the
polynomial in Eq.~(\ref{eq:P(x)}) as follows:

\begin{equation}
  \label{eq:P(x)farfromextremal}
  P(x)
  =
  |x -(b\alpha+ic\alpha^{1/4})|^{2}
  (x+d\alpha^{1/4}) (x-e\alpha^{1/4})
  [x-(x_{+}+f\alpha) ][x-(x_{-}+g\alpha) ]
  +\mathcal{O}(\alpha^{2})\, ,
\end{equation}

\noindent
with $x_{\pm} = 1\pm\sqrt{1-q^{2}}$. Again, comparing with  Eq.~(\ref{eq:P(x)})
we find

\begin{eqnarray}
  \label{eq:f}
  2f
  & = &
        -\frac{1}{\sqrt{1-q^{2}}} \left[\frac{9}{40}-\frac{21}{20 q^{2}} +\frac{4}{5q^{4}}
        \right]
        -\frac{13}{20 q^{2}}+\frac{4}{5q^{4}}+\mathcal{O}(\alpha)\, .
\end{eqnarray}

We have checked numerically that


\begin{equation}
  \label{eq:eventhorizon}
  \rho_{\rm h}
  =
  r_{+}
  +\frac{\alpha'}{M}\left\{
     -\frac{13}{20}(M/p)^{2}+\frac{8}{5}(M/p)^{4}
    -\frac{1}{\sqrt{1-\tfrac{1}{2}(p/M)^{2}}}
    \left[\frac{9}{80}-\frac{21}{20}(M/p)^{2} +\frac{8}{5}(M/p)^{4} \right]
  \right\}\, ,
\end{equation}

\noindent
gives a very good approximation to the position of the event horizon
to first order in $\alpha'$, even close to extremality. The position
of the inner horizon is given by

\begin{equation}
  \rho_{\rm -}
  =
  r_{-}
    +\frac{\alpha'}{M}\left\{
     -\frac{13}{20}(M/p)^{2}+\frac{8}{5}(M/p)^{4}
    +\frac{1}{\sqrt{1-\tfrac{1}{2}(p/M)^{2}}}
    \left[\frac{9}{80}-\frac{21}{20}(M/p)^{2} +\frac{8}{5}(M/p)^{4} \right]
  \right\}\, ,
\end{equation}

\noindent
but it is only good for small (but larger than 1) values of $\sqrt{2}(M/p)$.

In the near-extremality regime the square root term becomes imaginary before
extremality is reached. Therefore, we must make a different Ansatz for the
polynomial Eq.~(\ref{eq:P(x)}):

\begin{equation}
  \label{eq:P(x)nearextremal}
  P(x)
  =
  |x -(b\alpha+ic\alpha^{1/4})|^{2}
  (x+d\alpha^{1/4}) (x-e\alpha^{1/4})
  [x^{2} -2(1+h\alpha)x +(1+j\alpha)q^{2} ]
  +\mathcal{O}(\alpha^{2})\, ,
\end{equation}

\noindent
where $h$ and $j$ are two additional real constants which are found to have the
values

\begin{subequations}
  \begin{align}
    h
    & =
      -\frac{13}{40q^{2}} +\frac{2}{5q^{4}}\, ,
    \\
    & \nonumber \\
    j
    & =
      \frac{9}{40q^{2}} -\frac{17}{10q^{4}} +\frac{8}{5q^{6}}\, .
  \end{align}
\end{subequations}

The two roots corresponding to the horizons are

\begin{equation}
\tilde{x}_{\pm} = 1+h\alpha \pm \sqrt{1-q^{2} +(2h -q^{2}j)\alpha}\, ,
\end{equation}

\noindent
and, parametrizing $q$ near extremality by $q=1+\delta\alpha$ and replacing
$h$ and $j$ by their values, given above, we get

\begin{equation}
  \label{eq:rhopmnearextremal}
  \rho_{\pm}
  =
  M\pm  \sqrt{2\alpha'}\,
\sqrt{\frac{1}{80} -\delta} +\frac{3}{40 M}\alpha'
+\mathcal{O}(\alpha^{\prime\,  3/2}) \, .
\end{equation}

The extremal limit is $\delta=1/80$, and, there, we have

\begin{equation}
\rho_{\rm h\, ext} = M\left(1+\frac{3}{40}\frac{\alpha'}{M^{2}} \right)\, .
\end{equation}

Close to that limit, replacing $\delta \alpha'$ by its value $M(p/\sqrt{2}-M)$
in Eq.~(\ref{eq:rhopmnearextremal}), we find that the horizon is placed at

\begin{equation}
  \label{eq:rhohnearextremal0}
  \rho_{\rm h\, nearext}
  =
  M\left\{
    1+\sqrt{2}\sqrt{1+\frac{1}{80}\frac{\alpha'}{M^{2}} -\frac{p}{\sqrt{2}M}}
    +\frac{3}{40}\frac{\alpha'}{M^{2}}
    \right\}\, .
\end{equation}
\noindent

Nevertheless, it is useful to rewrite this formula explicitly in terms of the small quantity $M-M_{\rm ext}=M-\frac{p}{\sqrt{2}}+\frac{\sqrt{2}\alpha'}{80 p}$, in whose case it reads

\begin{equation}
  \label{eq:rhohnearextremal}
  \rho_{\rm h\, nearext}
  =\frac{p}{\sqrt{2}}+\frac{\sqrt{2}\alpha'}{16 p}+2^{1/4}p^{1/2}\sqrt{M-M_{\rm ext}}+(M-M_{\rm ext})+\ldots
\end{equation}

\section{Temperature}
\label{sec-temperature}

The Hawking temperature is related to the surface gravity by the famous
formula

\begin{equation}
T=\frac{\kappa}{2\pi}\, ,  
\end{equation}

\noindent
while the surface gravity of the event horizon of a static, spherically
symmetric black hole, located at $\rho=\rho_{\rm h}$, is given by

\begin{equation}
  \kappa
  =
  \tfrac{1}{2}\left[ \frac{1}{\sqrt{-g_{tt}g_{\rho\rho}}}\frac{dg_{tt}}{d\rho}\right]_{\rho_{\rm h}}\, .
\end{equation}

In terms of the variable $x\equiv \rho/M$ and taking into account that the
event horizon corresponds to the outermost first-order zero of the function
$f$ in Eq.~(\ref{eq:correcterdmetric3}), we find the following expression for
the surface gravity

\begin{equation}
  \label{eq:temperatureformula}
  T
  =
  \frac{1}{4\pi M}\frac{N(x_{\rm h})}{x^{6}_{\rm h}}
  \left[ \frac{P(x)}{x-x_{\rm h}}\right]_{x_{\rm h}}\, ,
\end{equation}

\noindent
where $P(x)$ is the polynomial defined in Eq.~(\ref{eq:P(x)}) if we are far
from the extremal limit. In that regime, the polynomial can be written in the
form Eq.~(\ref{eq:P(x)farfromextremal}) and $x_{\rm h}=x_{+}+f\alpha$, and,
therefore, we just have to evaluate at $x=x_{\rm h}$, to first order in
$\alpha$, the fifth-order polynomial

\begin{equation}
  \label{eq:P5(x)farfromextremal}
  \begin{aligned}
    \frac{P(x)}{x-x_{\rm h}}
    & =
    |x -(b\alpha+ic\alpha^{1/4})|^{2} (x+d\alpha^{1/4})
    (x-e\alpha^{1/4}) [x-(x_{-}+g\alpha) ] +\mathcal{O}(\alpha^{2})
    \\
    & \\
    & =
    \left[ x^{4}
      -\left(\frac{13}{20q^{2}}-\frac{4}{5q^{4}}\right)\alpha x^{3}
      -\left(\frac{9}{40}-\frac{2}{5q^{2}} \right)\alpha x^{2}
      +\frac{1}{5}\alpha x -\frac{11}{40}\alpha q^{2} \right]
    [x-(x_{-}+g\alpha) ]
    \\
    & \\
    & +\mathcal{O}(\alpha^{2})\, .
  \end{aligned}
\end{equation}

At first order, recognizing

\begin{equation}
  f+g
  =
  -\frac{13}{20q^{2}}+\frac{4}{5q^{4}}\, ,
\end{equation}

\noindent
we get

\begin{equation}
  \begin{aligned}
    \left[ \frac{P(x)}{x-x_{\rm h}}\right]_{x_{\rm h}}
    & =
    x_{+}^{4}(x_{+}-x_{-}) +\alpha \left\{
      x_{+}^{4}(f-g) + (x_{+}-x_{-})\left[  (5f+g) x_{+}^{3}
        \right.\right.
      \\
      & \\
      &
      \left.\left.
      -\left(\frac{9}{40}-\frac{2}{5q^{2}} \right)x_{+}^{2}
      +\frac{1}{5}x_{+} -\frac{11}{40}q^{2} \right]
      \right\} +\mathcal{O}(\alpha^{2})\, .
  \end{aligned}
\end{equation}

Plugging this result into Eq.~(\ref{eq:temperatureformula}) and operating 
we can write the temperature in the form

\begin{equation}
    T
     =
    T^{(0)}
 \left\{1   - \frac{\alpha}{x_{+}^{4}}
    \left[
      \frac{(g-f)}{x_{+}-x_{-}} q^{2}x_{+}^{2}
      +\left(\frac{9}{40}-\frac{2}{5q^{2}} \right)x_{+}^{2}
      -\frac{1}{5}x_{+} +\frac{3}{20}q^{2}
      \right]
    \right\}
    +\mathcal{O}(\alpha^{2})\, ,
\end{equation}

\noindent
where

\begin{equation}
  \label{eq:T0}
  T^{(0)}
  =
  \frac{(x_{+}-x_{-})}{4\pi M x_{+}^{2}}
  =
\frac{\sqrt{M^2-\frac{p^2}{2}}}{2\pi
  \left(M+\sqrt{M^2-\frac{p^2}{2}}\right)^2}\, ,
\end{equation}

\noindent
is the temperature of the uncorrected RN black hole. Operating with the actual
values of $f$ and $g$, and making use of the definitions



\begin{equation}
  q=p/(\sqrt{2}M)\, ,
  \hspace{.5cm}
  x_{\pm} = 1\pm \sqrt{1-q^{2}}\, ,
  \hspace{.5cm}
  \alpha
  =
  \alpha'/M^{2}\, ,
\end{equation}

\noindent
we obtain our final expression for the temperature,
\begin{figure}[t!]
\centering
\includegraphics[width=0.7\textwidth]{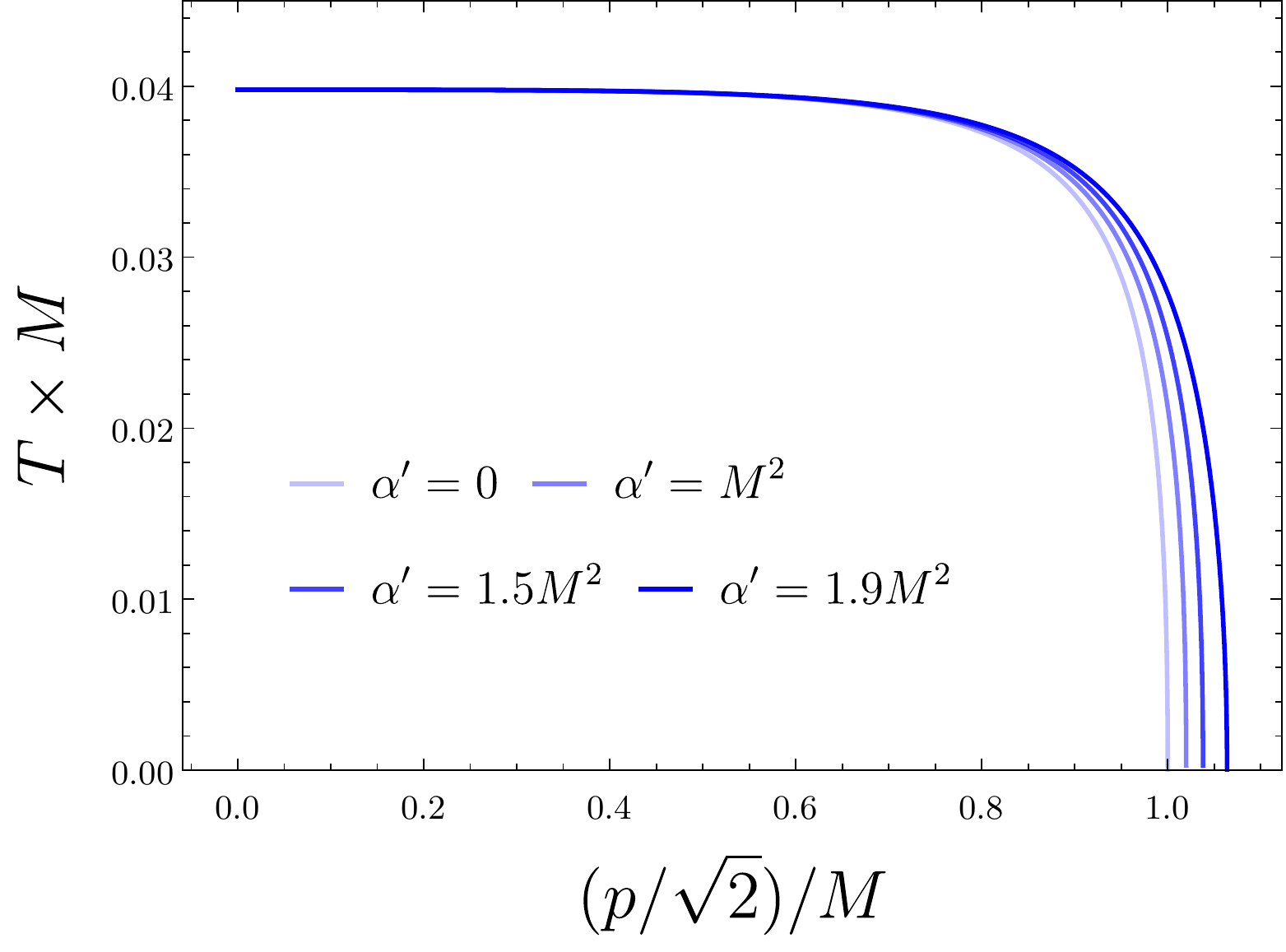}
\caption{\small Temperature of the $\alpha'$-corrected Reissner-Nordstr\"on black holes. We show $T$ as a function of the charge rescaling all quantities in terms of the mass $M$. The corrections are more relevant for large charges and we see that extremality is reached for $p/\sqrt{2}>M$. }
\label{fig:T}
\end{figure}

\begin{equation}\label{eq:temperature}
  T
  =
  \frac{\sqrt{M^2-\frac{p^2}{2}}}{2\pi
  \left(M+\sqrt{M^2-\frac{p^2}{2}}\right)^2}
+\frac{\alpha'  \left(M+3 \sqrt{M^2-\frac{p^2}{2}}\right) \left(M-\sqrt{M^2-\frac{p^2}{2}}\right)^2}{160 \pi  \sqrt{M^2-\frac{p^2}{2}} \left(M+\sqrt{M^2-\frac{p^2}{2}}\right)^5}+\mathcal{O}(\alpha'^2)\, .
\end{equation}

\noindent
This expression diverges for $M\rightarrow p/\sqrt{2}$, but this is simply
indicating that the approximation implied in (\ref{eq:P5(x)farfromextremal})
is no longer valid. Instead, in the near-extremal limit it is straightforward
to obtain



\begin{equation}
\label{eq:Tnearextremal}
T
=
\frac{1}{\pi p^2 }\left[
2^{1/4}p^{1/2}\sqrt{M-M_{\rm ext}}+4(M-M_{\rm ext})+\ldots\right]\, ,
\end{equation}
\noindent
where we recall that $M_{\rm ext}=\frac{p}{\sqrt{2}}-\frac{\sqrt{2}\alpha'}{80 p}+\mathcal{O}(\alpha'^2)$. Thus the
temperature vanishes in the limit $M\rightarrow M_{\rm ext}$, as it should. 
We note that in the expression above all the corrections enter implicitly through
the shift in the extremal mass, for formally the expansion in powers of $M-M_{\rm ext}$ is the
same as in the RN black hole. We observe that, near extremality, the
corrections to the temperature of a solution of fixed mass are of order $\alpha'^{1/2}$. In particular,
the solution with $M=p/\sqrt{2}$, which at zeroth order corresponds to the
extremal case, possesses a non-vanishing temperature
$T=\alpha'^{1/2}/(4\pi\sqrt{10} M^2)$. The complete profile of the temperature
as a function of the charge is shown in Fig.~\ref{fig:T} for few values of
$\alpha'/M^2$.

\section{Entropy}
\label{sec-entropy}

In order to compute the entropy of this black hole, it is necessary to
take into account the presence of higher-curvature terms in the
action.  Wald's entropy formula \cite{Wald:1993nt,Iyer:1994ys} takes
into account the possible presence of these terms and yields an entropy
that satisfies the first law of black-hole thermodynamics. However,
this formula was derived under the assumption that all the fields in
the theory are tensors. This is very restrictive, as the only physical
fields in our current description of Nature which are tensors, apart
from the metric, are scalars. Therefore, strictly speaking, it has not
been proven that Wald's formula can be applied even to the
Einstein-Maxwell theory, since the Maxwell field is not a tensor
field, but a connection. It is also unclear whether Wald's
formula can be applied to theories with fields with any kind of gauge
freedom, either. This is true even for General Relativity itself when
it is formulated in terms of a Vierbein! Fortunately, Jacobson and Moh
showed in Ref.~\cite{Jacobson:2015uqa} that, once the subtleties
associated to the (``induced'' or ``compensating'') local Lorentz
transformations that the Vierbein suffers when one acts on it with a
diffeomorphism are taking into account, Wald's formula can be applied
essentially unchanged.

The Heterotic Superstring effective action, reviewed in
Appendix~\ref{sec-heteroticalpha}, is a much more complicated beast,
though. To start with, it has to be formulated, necessarily, in terms
of a Zehnbein, in order to include spinor fields. One can deal with
both of them in the same way as Jacobson and Moh dealt with the
Vierbein in 4 dimensions: using the Lie-Lorentz
derivative.\footnote{See, for instance, Ref.~\cite{Ortin:2002qb} and
  references therein.}  Then, (most likely) one can prove that the
black-hole entropy is given by Wald's formula once again. However, the
action also includes Yang-Mills gauge fields which do not just occur
via the gauge-covariant Yang-Mills field strength but also via the
Chern-Simons 3-form Eq.~(\ref{eq:oYM}), which transforms in a
completely different way. Actually, the same happens to the Zehnbein:
it also occurs in the action via the Chern-Simons 3-form of the spin
connection 1-form Eq.~(\ref{eq:oL0}). This does not mean that the
action is not gauge- or Lorentz-invariant, because these terms only
occur via the 3-form field strength in Eq.~(\ref{eq:H1}) which is gauge-
and local-Lorentz invariant thanks to the very special way in which
the Kalb-Ramond 2-form behaves under gauge and local-Lorentz
transformations (the so-called Nicolai-Townsend
transformations). Taking all this into account, it has been shown in
Ref.~\cite{Edelstein:2019wzg} that Wald's formula also applies to the
Heterotic Superstring effective
action,\footnote{Ref.~\cite{Edelstein:2019wzg} deals with a family of
  actions which is, in certain respects, more general than the
  Heterotic Superstring's but which do not include Yang-Mills gauge
  fields. However, there is no real difference between the behavior of
  gauge fields and local-Lorentz tensors or spinors and it is clear
  that the results obtained can be extended to include Yang-Mills
  fields straightforwardly.  On the other hand, in
  Ref.~\cite{Edelstein:2019wzg} it is assumed (but not directly
  proven) that a generalization of the Lie-Lorentz derivative can be
  constructed. This point clearly deserves further investigation.}
justifying the results obtained in
Refs.~\cite{Cano:2018qev,Cano:2018brq}.

Wald's formula for the black-hole entropy can be written in the
form

\begin{equation}
  \label{eq:entropyd4}
S
  =
  -2\pi\int_\Sigma d^{2}x\sqrt{|h|}
  \frac{\partial \mathcal{L}}{\partial R_{abcd}}
 \epsilon_{ab}\epsilon_{cd}\, .
\end{equation}

\noindent
where $|h|$ is the absolute value of the determinant of the metric
induced over the event horizon, $\epsilon^{ab}$ is the event horizon's
binormal normalized so that $\epsilon_{ab}\epsilon^{ab}=-2$ and
$R_{abcd}$ is the Riemann tensor. We will work in the (``modified'')
Einstein-frame metric. Then, it can be shown \cite{kn:????} that the
partial derivative of the Heterotic Superstring effective action
compactified on a trivial T$^{5}$ and then on S$^{1}$ with respect to
the Riemann tensor of that conformal frame is given in terms of
4-dimensional objects by

\begin{equation}
  \label{eq:dLdR}
  \begin{aligned}
    \frac{\partial{\mathcal{L}}}{\partial {R}_{abcd}}
    & =
    \frac{1}{16\pi G_{N}^{(4)} } \left\{ {g}^{ab,\, cd}
      -\frac{\alpha'}{8} \left[e^{-4(\phi-\phi_{\infty})}{H}^{(0)\,
          abg} \left(\omega_{g}{}^{cd}+2\Sigma_{g}{}^{cd}\right)
      \right.\right.
    \\
    & \\
    & \left. \left.
        +e^{-2(\phi-\phi_{\infty})}
        \left(
          -2\tilde{R}_{(-)}^{(0)\, abcd} +K^{(-)\, [a|c}K^{(-)\, |b]d}
          +K^{(+)\, ab}K^{(+)\, cd}
        \right)
      \right]
    \right\}\, ,
    \\
     \end{aligned}
\end{equation}

\noindent
where

\begin{eqnarray}
  H^{(0)}{}_{\mu\nu\rho}
  & \equiv &
             3\partial_{[\mu}B^{(0)}{}_{\nu\rho]}
      -\tfrac{3}{2}A_{[\mu}G^{(0)}{}_{\nu\rho]}
             -\tfrac{3}{2}B^{(0)}{}_{[\mu}F_{\nu\rho]}\, ,
  \\
  & & \nonumber \\
  K^{(\pm)}{}_{\mu\nu}
  & \equiv & 
             kF_{\mu\nu}\pm k^{-1}G^{(0)}{}_{\mu\nu}\, ,
  \\
  & & \nonumber \\
  \Sigma_{\mu}{}^{a}{}_{b}
  & \equiv &
             \Delta_{\mu}{}^{a}{}_{b} -\tfrac{1}{2}H^{(0)}{}_{\mu}{}^{a}{}_{b}\, ,
\\
  & & \nonumber \\
\Delta_{\mu\, ab}
    & \equiv &
    -\partial_{[\mu}\phi\eta_{b]c}
    +{e}_{[c|\, \mu}{e}_{|b]}{}^{\nu}\partial_{\nu}\phi
               -{e}_{[c|}{}^{\nu}{e}_{b\, |\mu]}\partial_{\nu}\phi\, ,             \\
  & & \nonumber \\
  \tilde{\Omega}_{(-)\, \mu}^{(0)}{}^{a}{}_{b}
  & \equiv &
\omega_{\mu}{}^{a}{}_{b}
 +\Sigma_{\mu}{}^{a}{}_{b}\, .
\end{eqnarray}

\noindent
Here $A_{\mu}$ is the KK vector field and
$F_{\mu\nu}=2\partial_{[\mu}A_{\nu]}$ its field strength,
$B^{(0)}{}_{\mu}$ is the winding vector field at zeroth-order in
$\alpha'$ and $G^{(0)}{}_{\mu\nu}=2\partial_{[\mu}B^{(0)}{}_{\nu]}$
its field strength, $B^{(0)}{}_{\mu\nu}$ is the Kalb-Ramod 2-form at
zeroth order in $\alpha'$ and $H^{(0)}{}_{\mu\nu\rho}$ its
gauge-invariant field strength expressed in a manifestly
T-duality-invariant form. Furthermore,
$\tilde{R}^{(0)}_{(-)\, \mu\nu}{}^{a}{}_{b}$ is the curvature 2-form
of the connection $\tilde{\Omega}_{(-)\, \mu}^{(0)}{}^{a}{}_{b}$,
which differs from the usual torsionful spin connection
$\Omega_{(-)\, \mu}^{(0)}{}^{a}{}_{b}$ by the dilaton-dependent
$\Delta_{\mu}{}^{a}{}_{b}$ contribution which arises in the Weyl
rescaling from the string to the Einstein frame.

The uncorrected RN black hole has $k=1, F_{\mu\nu}=H^{(0)}{}_{\mu\nu\rho}=0$
and $\phi=\phi_{\infty}$ at zeroth order in $\alpha'$, which means that
$\tilde{R}^{(0)}_{(-)\, \mu\nu}{}^{a}{}_{b}= R^{(0)}_{(-)\,
  \mu\nu}{}^{a}{}_{b} =R^{(0)}_{\mu\nu}{}^{a}{}_{b}$, the Riemann curvature of
the original, uncorrected, RN black hole. Wald's formula in $G_{N}^{(4)}=1$
units takes the form

\begin{equation}
  \label{eq:Waldsformulafortheseblackholes}
 \begin{aligned}
   S & = -\frac{1}{8}\int_{\Sigma} d^{2}x\sqrt{|h|}
   \epsilon_{ab}\epsilon_{cd} \left\{ {g}^{ab,\, cd} \right.
   \\
   & \\
   & \left.  -\frac{\alpha'}{8} \left[ -2R_{(-)}^{(0)\, abcd}
       +G^{(0)\, [a|c}G^{(0)\, |b]d} +G^{(0)\, ab}G^{(0)\, cd} \right]
   \right\}
   \\
   & \\
   & = -\frac{1}{8}\int_{\Sigma} d^{2}x\sqrt{|h|} \left\{-2
     +\alpha'\left[R^{(0)\, 0101} -\tfrac{3}{4}\left(G^{(0)\,
           01}\right)^{2} \right] \right\}
   \\
   & \\
   & = \frac{A_{\rm h}}{4}
   -\alpha'\left[\tfrac{1}{2}(a^{2})_{\rm h}''
     -\frac{3p^{2}}{4\rho_{\rm h}^{4}}
   \right]\frac{A^{(0)}_{\rm h}}{8}
 \\
 & \\
 & = \pi\rho^{2}_{\rm h}\left\{1
 +\alpha'\left[\frac{M}{\rho^{3}_{\rm h}} -\frac{3p^{2}}{8\rho_{\rm h}^{4}}
 \right]\right\}\, ,
     \end{aligned}
\end{equation}

\noindent
where $\rho_{\rm h}$ is the radius of the event horizon, and $A_{\rm h}$ is
the area of the event horizon, $4\pi \rho^{2}_{\rm h}$.

This formula, which we rewrite here for the sake of convenience,

\begin{equation}
  \label{eq:entropy}
  S
  =
  \pi\rho^{2}_{\rm h}\left\{1
 +\alpha'\left[\frac{M}{\rho^{3}_{\rm h}} -\frac{3p^{2}}{8\rho_{\rm h}^{4}}
 \right]\right\}\, ,  
\end{equation}

\noindent
is one of the main results of this paper, but we must test it against the
temperature computed in Section~\ref{sec-temperature}.

Far from the extremal limit we can use the value of the radius of the horizon $\rho_h$ given
in Eq.~\eqref{eq:eventhorizon}, and after some simplifications we arrive at the following result for the entropy,
\begin{equation}
  \label{eq:Sfarextremal}
  S
  =
  \pi\left[ \left(M+\sqrt{M^2-\frac{p^2}{2}}\right)^2+\frac{ \alpha'  \left(18 M \sqrt{M^2-\frac{p^2}{2}}+21 \left(M^2-\frac{p^2}{2}\right)+M^2\right)}{40  \sqrt{M^2-\frac{p^2}{2}} \left(\sqrt{M^2-\frac{p^2}{2}}+M\right)}\right]+\mathcal{O}(\alpha'^2)\, .
\end{equation}

\noindent
Now, in a thermodynamic system with energy $M$ and entropy $S$, the
temperature is defined through the standard relation

\begin{equation}
\frac{\partial S}{\partial M}=\frac{1}{T}\, ,
\end{equation}

\noindent
and in the case of black holes this temperature should coincide with the
Hawking's one on account of the first law of black hole mechanics. By deriving Eq.~(\ref{eq:Sfarextremal}) with respect to $M$, it is easy to
check that the thermodynamic temperature agrees with Hawking's temperature in
Eq.~(\ref{eq:temperature}) at first order in $\alpha'$, which constitutes a
strong consistency test of our computations.












In the near-extremal limit, according to Eq.~(\ref{eq:rhohnearextremal}), which
we reproduce here for the sake of convenience, we have

\begin{equation}
  \rho_{\rm h\, nearext}
  =\frac{p}{\sqrt{2}}+\frac{\sqrt{2}\alpha'}{16 p}+2^{1/4}p^{1/2}\sqrt{M-M_{\rm ext}}+(M-M_{\rm ext})+\ldots
\end{equation}

\noindent
and, substituting this value in Eq.~(\ref{eq:entropy}) we get


\begin{equation}
  \label{eq:Snearextremal}
  S/\pi
  =
 \frac{p^2}{2}+\frac{3\alpha'}{8}+2^{3/4}p^{3/2}\sqrt{M-M_{\rm ext}}+\sqrt{8}p (M-M_{\rm ext})+\ldots\, ,
\end{equation}

\noindent
It is straightforward to check that the entropy and temperature in the
near-extremal regime, given by Eqs.~(\ref{eq:Snearextremal}) and
(\ref{eq:Tnearextremal}) also satisfy the thermodynamic relation
$\partial S/\partial M = T^{-1}$. 
If we take the extremal limit in this expression, $M\rightarrow M_{\rm ext}$, we observe that the entropy gets an $\mathcal{O}(\alpha')$ correction

\begin{equation}
  S
  =
  S^{(0)}_{\rm ext}+\frac{3\pi}{8}\alpha'\, ,
  \,\,\,\,\,\,
  \text{where}
  \,\,\,\,\,\,
  S^{(0)}_{\rm ext}
  =
\pi p^{2}/2\, ,
\end{equation}
\noindent
However, this expression should not be trusted due to the presence of logarithmic divergences of some of the fields (which are generically found at the Cauchy horizon) at the event horizon. Indeed, from \eqref{eq:dLdR} it is manifest that the dilaton divergence would produce an infinite correction to the entropy, which is meaningless. Hence, the analysis presented here is only valid for non-extremal configurations. In the extremal limit, it seems the black hole becomes singular after the higher-curvature corrections are incorporated so it makes no sense to attribute a value to its entropy.

Near-extremality, the corrections to the entropy are of order $\alpha'^{1/2}$ as in Ref.~\cite{Loges:2019jzs}. In particular, for the solution with $M=p/\sqrt{2}\equiv M_{\rm ext}^{(0)}$ we find
\begin{equation}\label{eq:Smext0}
S\Big|_{M=M_{\rm ext}^{(0)}}=\pi\left[\frac{p^2}{2}+\frac{p \alpha'^{1/2}}{2\sqrt{5}}+\mathcal{O}(\alpha')\right]\, .
\end{equation}

\section{Discussion}
\label{sec-discussion}

In this paper we have computed the first-order in $\alpha'$ corrections to a
dyonic Reissner-Nordstr\"om black hole explicitly embedded in the Heterotic String Theory. To the best
of our knowledge, this is the first explicit example of a non-extremal Reissner-Nordstr\"om solution containing all of the $\alpha'$-corrections. In the extremal limit, we have seen that the charge-to-mass ratio of the solution is
positively corrected

\begin{equation}
  \frac{p/\sqrt{2}}{M}\bigg|_{\rm ext}
  =
  1+\frac{\alpha'}{80 M^2}+\mathcal{O}(\alpha'^2)\, ,
\end{equation}

\noindent
in agreement with the mild form of the weak gravity conjecture. Nevertheless, since it seems that the black hole becomes singular in that limit, this only provides some sort of indirect signal in favour of the conjecture. In the context of HST, the first example of regular extremal black hole with corrections to the charge-to-mass ratio in agreement with the WGC was recently given in Ref.~\cite{Cano:2019oma}.\footnote{The GHS solution \cite{Gibbons:1982ih,Gibbons:1987ps, Garfinkle:1990qj}, whose corrections where obtained in Ref.~\cite{Kats:2006xp}, does not describe a black hole in the extremal limit.} 

We have also computed the corrections to the temperature and to the entropy of
these black holes --- see (\ref{eq:temperature}) and
(\ref{eq:Sfarextremal}). The temperature is straightforwardly computed from
the surface gravity of the horizon, but the calculation of the entropy through
the evaluation of Wald's formula presents some subtleties associated to the
presence of Chern-Simons terms in the 10-dimensional action. Those subtleties
can be handled with the methods used in Refs.~\cite{Cano:2018qev,Cano:2018brq}
but it is most reassuring to see that they completely disappear for these
black-hole solutions when the 10-dimensional action is compactified as in
Ref.~ \cite{kn:????} and Wald's formula takes the explicitly gauge-invariant
form Eq.~(\ref{eq:Waldsformulafortheseblackholes}).\footnote{For more general
  solutions one has to use Eq.~(\ref{eq:dLdR}), though. This expression
  contains explicit contributions from the spin connection which are not
  manifestily invariant under local Lorentz transformations and, at this
  point, it is not clear if they give non-trivial contributions to the
  entropy.} Another possibility to compute the entropy would be to rewrite the
HST action in a gauge-invariant manner without performing the dimensional
reduction as in \cite{Faedo:2019xii}. We have checked that the application of this
method produces the same result \eqref{eq:Sfarextremal}. As a highly
non-trivial check of our computations, we have shown that the thermodynamic
relation $\partial S/\partial M=1/T$ holds at order $\alpha'$.

We have found that the entropy shift is always positive for this family of solutions. In previous works in the literature, it
has been claimed that the positivity of the corrections to the
entropy imply a positive correction to the charge-to-mass ratio at extremality \cite{Cheung:2018cwt, Cheung:2019cwi, Goon:2019faz}. On the contrary, this claim has been disputed by the counterexample presented in Ref.~\cite{Cano:2019oma}, in which $\Delta S >0$ but $\Delta (q/M)=0$. It is interesting to ask
what the situation is here, since we have a non-extremal solution at our
disposal and we can perform a more detailed analysis. In Ref.~\cite{Goon:2019faz} the shifts are claimed to
satisfy a universal relation,

\begin{equation}
\label{eq:shiftrelations}
\Delta M_{\text{ext}} = - T_0 (M, \vec{Q}) \Delta S (M, \vec{Q}) \vert_{M \approx M^{(0)}_{\text{ext}}} \, .
\end{equation} 

\noindent
Here, $\Delta M_{\text{ext}} $ is the change in the energy of the solution at zero temperature, while $T_0 (M, \vec{Q})$ and $\Delta S (M, \vec{Q})$ are, respectively, the unperturbed temperature and the shift in the entropy for fixed values of the mass and charges. As $T_0(M, \vec{Q})$ is parametrically small for $M \rightarrow M^{(0)}_{\text{ext}}$, one sees that, whenever $\Delta M_{\text{ext}} \neq 0$, the expression for $\Delta S (M, \vec{Q})$ that must be used in \eqref{eq:shiftrelations} becomes divergent as $M \rightarrow M^{(0)}_{\text{ext}}$, so it cannot really correspond to the correction to the entropy for this value of the mass, which should be finite. For this reason, according to the prescription given in Ref.~\cite{Goon:2019faz}, the right hand side of \eqref{eq:shiftrelations} must be evaluated taking $M$ to be slightly larger than the unperturbed extremal limit, which is denoted as $M \approx M^{(0)}_{\text{ext}}$, defined such that the corrections to the temperature at fixed mass and charges are subdominant. In the particular case we study in this article, this could be expressed as follows,
\begin{equation}
\alpha'/p<<M-\frac{p}{\sqrt{2}}<<p\, .
\end{equation}
In this regime, the right hand side of \eqref{eq:shiftrelations}, computed using the perturbative correction to the entropy given in expression \eqref{eq:Sfarextremal}  and the uncorrected temperature in \eqref{eq:T0}, yields the right value of $\Delta M_{\text{ext}}$ for our solution at the order we are working.

However, it would be convenient to have an expression similar to \eqref{eq:shiftrelations} in which the ambiguity in the value of evaluation of $M$  is eliminated. For $M-\frac{p}{\sqrt{2}} \sim \mathcal{O}(\alpha'/p)$, Eq.~\eqref{eq:shiftrelations} cannot be correct because, as we said, it would require the entropy to be divergent. Nevertheless, in our solution the correction to the entropy remains finite in that regime, which we might call the ``very near-extremal" regime.  In particular, by explicit evaluation we find the following relation for our solution:
\begin{equation}
\label{eq:shiftrelationsgood}
\Delta M_{\text{ext}}=-\frac{1}{2} T(M_{\text{ext}}^{(0)},\vec{Q}) \Delta S(M_{\text{ext}}^{(0)},\vec{Q})\, ,
\end{equation}
where now $T(M_{\text{ext}}^{(0)},\vec{Q})$ is the actual (corrected) value of the temperature --- see \eqref{eq:Tnearextremal} --- for the solution with $M=M_{\text{ext}}^{(0)}$, while $\Delta S(M_{\text{ext}}^{(0)},\vec{Q})$ is the correction to the entropy of the extremal black hole for fixed mass and charges, which is of order $\alpha'^{1/2}$ as shown in \eqref{eq:Smext0}.
We can see, through a very simple argument, that this formula probably holds in general. 
Near-extremality, the entropy will generically have the following expansion as a function of the mass (keeping the charges constant), 
\begin{equation}
\label{eq:Sgenexp}
S = S_{\text{ext}}(\vec{Q})+ k(\vec{Q}) \sqrt{M-M_{\text{ext}}} + \dots \, ,
\end{equation}
\noindent
for some function of the charges $k(\vec{Q})$ and where $S_{\text{ext}}(\vec{Q})$ is the entropy at extremality (containing the corresponding corrections). The fact that the first term comes with a fractional power of $(M-M_{\text{ext}})$ is consequence of the first law of thermodynamics, as $\partial S/\partial M = T^{-1}$ diverges in the zero temperature limit.
Then, taking the derivative of \eqref{eq:Sgenexp}, using the first law and evaluating at $M=M_{\text{ext}}^{(0)}$ (which is consistent only if $M_{\text{ext}}^{(0)}\geq M_{\text{ext}}$), it is straightforward to get
\begin{equation}
\label{eq:shiftrelationsgood2}
\Delta M_{\text{ext}}=-\frac{1}{2} T(M_{\text{ext}}^{(0)}) \left[ S(M_{\text{ext}}^{(0)},\vec{Q})-S_{\text{ext}}(\vec{Q}) \right]  \, ,
\end{equation}
and then it is easy to note that, to leading order $S(M_{\text{ext}}^{(0)},\vec{Q})-S_{\text{ext}}(\vec{Q})=\Delta S(M_{\text{ext}}^{(0)},\vec{Q})$, since the leading corrections to the entropy come from the term $\sqrt{M-M_{\text{ext}}}$ and are of order $\alpha'^{1/2}$. On the other hand, the corrections to the extremal entropy generically appear at first order in $\alpha'$ and they play no role in the relation \eqref{eq:shiftrelationsgood}. 

Equation \eqref{eq:shiftrelationsgood2} clarifies the relation between the perturbations to the entropy at fixed mass and charges and the shift to the charge-to-mass ratio at extremality. If $\Delta M_{\text{ext}} \neq 0$, a positive value of $\Delta S(M_{\text{ext}}^{(0)},\vec{Q})$ implies a positive correction to the charge-to-mass ratio. However, it is also possible to have $\Delta S(M_{\text{ext}}^{(0)},\vec{Q})>0$ and $\Delta M_{\text{ext}} = 0$, since in that case the relation \eqref{eq:shiftrelationsgood2} is trivially satisfied because no correction to the extremal mass implies $T(M_{\text{ext}}^{(0)})=0$. Hence, we conclude that the fact that the perturbation to the entropy at fixed mass is positive\footnote{The positivity of this variable is to be expected when the perturbation is due to the inclusion of higher-derivative operators motivated by the UV-completion of the effective theory --- see \cite{Cheung:2018cwt}.} does not imply the mild version of the Weak Gravity Conjecture. 
This observation clarifies the counterexample found in Ref.~\cite{Cano:2019oma}. 

One of the most important lessons we extract from the results we presented
here is that String Theory requires the activation of many additional fields
when higher-derivative corrections are taken into account. Thus, our staring
point was a dyonic Reissner-Nordstr\"om black hole, which is a solution of
Einstein-Maxwell theory. However, when that solution is embedded in the HST,
not only we get corrections to the metric and to the Maxwell field, but also
new fields acquire a non-trivial profile. In the case at hands, we activate
three scalars: the dilaton, an axion and a Kaluza-Klein scalar, and
a Kaluza-Klein vector field.

The exploration of constraints on the higher-derivative corrections
to simple models such as Einstein-Maxwell  or
Einstein-Maxwell-dilaton (EMD)\footnote{This model arises from the the
  effective theory of the Heterotic Superstring compactified on $T^{6}$
  ($\mathcal{N}=4,d=4$ supergravity) after several truncations are made
  \cite{Gibbons:1982ih}. The consistency of those truncations is ensured by
  the fact that they are performed in the equations of motion. The action of
  the EMD model leads to the truncated equations of motion. The black-hole
  solutions of this model were first found in Ref.~\cite{Gibbons:1982ih} and
  rederived later on in Refs.~\cite{Gibbons:1987ps, Garfinkle:1990qj}. Further
  truncation to the Einstein-Maxwell model can be achieved by
  constraining the form of the Maxwell field, which has to be dyonic with
  equal electric and magnetic charges (or one has to introduce several Abelian
  vector fields with electric and magnetic charges). Thus, not all the
  solutions of the Einstein-Maxwell theory can be embedded in the Heterotic
  Superstring effective action because unequal electric and magnetic charges
  always generate a non-trivial scalar field.}
 theories inspired by \emph{quantum black hole} physics is currently attracting much attention \cite{Kats:2006xp,Cheung:2014ega,Charles:2017dbr,Cheung:2018cwt,Hamada:2018dde,Bellazzini:2019xts, Cheung:2019cwi,Loges:2019jzs,Goon:2019faz,Charles:2019qqt}. A recurrent assumption in these explorations is that no additional degrees of freedom are
activated at higher orders. In the light of the results presented here and in
previous literature \cite{Campbell:1990ai,Campbell:1990fu,Mignemi:1992pm,Cano:2019oma}, it is reasonable to wonder whether
this assumption can have significant consequences. The activation of
additional fields due to higher-derivative terms seems to be quite a generic feature of String Theory, and truncating the new fields might be
inconsistent in this context.

In our current analysis, the additional fields acquire a non-trivial profile
of order $\mathcal{O}(\alpha')$, which implies that they will backreact on the
geometry at order $\mathcal{O}(\alpha'^2)$. Thus, the additional degrees of
freedom do not play a role in the corrections to the entropy or to the
extremality bound at leading order in $\alpha'$, but they sure will do so at
$\mathcal{O}(\alpha'^2)$ and higher orders. Thus, the presence of new degrees
of freedom cannot be ignored in order to analyze, for instance, the positivity
of the corrections to the entropy beyond first order in the perturbative
expansion.  In fact, it would be interesting to obtain the
$\mathcal{O}(\alpha'^2)$ corrections to the solution we have studied, or to
the ones presented in \cite{Cano:2019oma}. This is perhaps a less challenging
task than it would appear, since no $\alpha'^2$ terms occur explicitly in the
HST effective action (they only appear implicitly through the iterative
definition of the 3-form field strength $\hat H$). Work in this direction is
alredy in progress \cite{kn:CMOR}.

\section*{Acknowledgments}

This work has been supported in part by the MCIU, AEI, FEDER (UE) grant
PGC2018-095205-B-I00, CONACyT grant 237351 and by the Spanish Research Agency (Agencia Estatal de
Investigaci\'on) through the grant IFT Centro de Excelencia Severo Ochoa
SEV-2016-0597.  PAC was mostly funded by Fundaci\'on la Caixa through a ``la Caixa - Severo Ochoa'' International pre-doctoral grant. In the last stages of this project PAC was supported by the KU Leuven grant ``Bijzonder Onderzoeksfonds C16/16/005 – Horizons in hoge-energie fysica". R.L. is thankful for warm hospitality by IFT. PFR is funded by the Alexander von Humboldt Foundation. TO wishes to thank M.M.~Fern\'andez for her 
permanent support.

\appendix

\section{The Heterotic Superstring effective action to
  \texorpdfstring{$\mathcal{O}(\alpha')$}{O(α')}}
\label{sec-heteroticalpha}

In order to describe the Heterotic Superstring effective action to
$\mathcal{O}(\alpha')$ as given in Ref.~\cite{Bergshoeff:1989de} (but
in the string frame), we start by defining the zeroth-order 3-form
field strength of the Kalb-Ramond 2-form $B$:

\begin{equation}
H^{(0)} \equiv dB\, ,  
\end{equation}

\noindent
and constructing with it the zeroth-order torsionful spin connections

\begin{equation}
{\Omega}^{(0)}_{(\pm)}{}^{{a}}{}_{{b}} 
=
{\omega}^{{a}}{}_{{b}}
\pm
\tfrac{1}{2}{H}^{(0)}_{{\mu}}{}^{{a}}{}_{{b}}dx^{{\mu}}\, ,
\end{equation}

\noindent
where ${\omega}^{{a}}{}_{{b}}$ is the Levi-Civita spin connection
1-form.\footnote{We follow the conventions of Ref.~\cite{Ortin:2015hya} for
  the spin connection and the curvature.}  With them we
define the zeroth-order Lorentz curvature 2-form and Chern-Simons 3-forms

\begin{eqnarray}
{R}^{(0)}_{(\pm)}{}^{{a}}{}_{{b}}
& = & 
d {\Omega}^{(0)}_{(\pm)}{}^{{a}}{}_{{b}}
- {\Omega}^{(0)}_{(\pm)}{}^{{a}}{}_{{c}}
\wedge  
{\Omega}^{(0)}_{(\pm)}{}^{{c}}{}_{{b}}\, ,
\\
  & & \nonumber \\
  \label{eq:oL0}
{\omega}^{{\rm L}\, (0)}_{(\pm)}
& = &  
d{\Omega}^{ (0)}_{(\pm)}{}^{{a}}{}_{{b}} \wedge 
{\Omega}^{ (0)}_{(\pm)}{}^{{b}}{}_{{a}} 
-\tfrac{2}{3}
{\Omega}^{ (0)}_{(\pm)}{}^{{a}}{}_{{b}} \wedge 
{\Omega}^{ (0)}_{(\pm)}{}^{{b}}{}_{{c}} \wedge
{\Omega}^{ (0)}_{(\pm)}{}^{{c}}{}_{{a}}\, .  
\end{eqnarray}

Next, we introduce the gauge fields. We will only activate a
$\mathrm{SU}(2)\times \mathrm{SU}(2)$ subgroup of the full gauge group
of the Heterotic Theory and we will denote by
$A^{A_{1,2}}$ ($A_{1,2}=1,2,3$) the components. The gauge field strength and
the Chern-Simons 3-form of each $\mathrm{SU}(2)$ factor are defined by

\begin{eqnarray}
{F}^{A}
& = & 
d{A}^{A}+\tfrac{1}{2}\epsilon^{ABC}{A}^{B}\wedge{A}^{C}\, , 
\\
  & & \nonumber \\
  \label{eq:oYM}
{\omega}^{\rm YM}
& = & 
dA^{A}\wedge {A}^{A}
+\tfrac{1}{3}\epsilon^{ABC}{A}^{A}\wedge{A}^{B}\wedge{A}^{C}\, .
\end{eqnarray}

Then, we are ready to define recursively 

\begin{eqnarray}
  \label{eq:H1}
H^{(1)}
& = & 
d{B}
+\frac{\alpha'}{4}\left({\omega}^{\rm YM}
+{\omega}^{{\rm L}\, (0)}_{(-)}\right)\, ,  
\nonumber \\
& & \nonumber \\
{\Omega}^{(1)}_{(\pm)}{}^{{a}}{}_{{b}} 
& = & 
{\omega}^{{a}}{}_{{b}}
\pm
\tfrac{1}{2}{H}^{(1)}_{{\mu}}{}^{{a}}{}_{{b}}dx^{{\mu}}\, ,
\nonumber \\
& & \nonumber \\
{R}^{(1)}_{(\pm)}{}^{{a}}{}_{{b}}
& = & 
d {\Omega}^{(1)}_{(\pm)}{}^{{a}}{}_{{b}}
- {\Omega}^{(1)}_{(\pm)}{}^{{a}}{}_{{c}}
\wedge  
{\Omega}^{(1)}_{(\pm)}{}^{{c}}{}_{{b}}\, ,
\nonumber  \\
& & \nonumber \\
{\omega}^{{\rm L}\, (1)}_{(\pm)}
& = & 
d{\Omega}^{(1)}_{(\pm)}{}^{{a}}{}_{{b}} \wedge 
{\Omega}^{(1)}_{(\pm)}{}^{{b}}{}_{{a}} 
-\tfrac{2}{3}
{\Omega}^{(1)}_{(\pm)}{}^{{a}}{}_{{b}} \wedge 
{\Omega}^{(1)}_{(\pm)}{}^{{b}}{}_{{c}} \wedge
{\Omega}^{(1)}_{(\pm)}{}^{{c}}{}_{{a}}\, .  
\nonumber \\
& & \nonumber \\
H^{(2)}
& = &  
d{B}
+\frac{\alpha'}{4}\left({\omega}^{\rm YM}
+{\omega}^{{\rm L}\, (1)}_{(-)}\right)\, ,  
\end{eqnarray}

\noindent
and so on.

In practice only $\Omega^{(0)}_{(\pm)},{R}^{(0)}_{(\pm)}, \omega^{{\rm L}\,
  (0)}_{(\pm)}, H^{(1)}$ will occur to the order we want to work at, but, often,
it is more convenient to work with the higher-order objects ignoring the terms of
higher order in $\alpha'$ when necessary. Thus we will suppress the $(n)$
upper indices from now on.

Finally, we define three ``$T$-tensors'' associated to the $\alpha'$
corrections

\begin{equation}
\label{eq:Ttensors}
\begin{array}{rcl}
{T}^{(4)}
& \equiv &
\dfrac{3\alpha'}{4}\left[
{F}^{A}\wedge{F}^{A}
+
{R}_{(-)}{}^{{a}}{}_{{b}}
\wedge
{R}_{(-)}{}^{{b}}{}_{{a}}
\right]\, ,
\\
& & \\ 
{T}^{(2)}{}_{{\mu}{\nu}}
& \equiv &
\dfrac{\alpha'}{4}\left[
{F}^{A}{}_{{\mu}{\rho}}{F}^{A}{}_{{\nu}}{}^{{\rho}} 
+
{R}_{(-)\, {\mu}{\rho}}{}^{{a}}{}_{{b}}
{R}_{(-)\, {\nu}}{}^{{\rho}\,  {b}}{}_{{a}}
\right]\, ,
\\
& & \\    
{T}^{(0)}
& \equiv &
{T}^{(2)\,\mu}{}_{{\mu}}\, .
\\
\end{array}
\end{equation}

In terms of all these objects, the Heterotic Superstring effective action in
the string frame and to first-order in $\alpha'$ can be written as

\begin{equation}
\label{heterotic}
{S}
=
\frac{g_{s}^{2}}{16\pi G_{N}^{(10)}}
\int d^{10}x\sqrt{|{g}|}\, 
e^{-2{\phi}}\, 
\left\{
{R} 
-4(\partial{\phi})^{2}
+\tfrac{1}{2\cdot 3!}{H}^{2}
-\tfrac{1}{2}T^{(0)}
\right\}\, ,
\end{equation}

\noindent
where $G_{N}^{(10)}$ is the 10-dimensional Newton constant, $\phi$ is the
dilaton field and the vacuum expectation value of $e^{\phi}$ is the Heterotic
Superstring coupling constant $g_{s}$. $R$ is the Ricci scalar of the
string-frame metric $g_{\mu\nu}$.

The derivation of the complete equations of motion is quite a complicated
challenge.  Following Ref.~\cite{Bergshoeff:1992cw}, we separate the
variations with respect to each field into those corresponding to occurrences
via ${\Omega}_{(-)}{}^{{a}}{}_{{b}}$, that we will call \textit{implicit}, and
the rest, that we will call \textit{explicit}:

\begin{eqnarray}
\delta S 
& = &  
\frac{\delta S}{\delta g_{\mu\nu}}\delta g_{\mu\nu}
+\frac{\delta S}{\delta B_{\mu\nu}}\delta B_{\mu\nu}
+\frac{\delta S}{\delta A^{A_{i}}{}_{\mu}}\delta A^{A_{i}}{}_{\mu}
+\frac{\delta S}{\delta \phi} \delta \phi
\nonumber \\
& & \nonumber \\
& = & 
\left.\frac{\delta S}{\delta g_{\mu\nu}}\right|_{\rm exp.}\delta g_{\mu\nu}
+\left.\frac{\delta S}{\delta B_{\mu\nu}}\right|_{\rm exp.}\delta B_{\mu\nu}
+\left.\frac{\delta S}{\delta A^{A_{i}}{}_{\mu}}\right|_{\rm exp.}
\delta A^{A_{i}}{}_{\mu}
+\frac{\delta S}{\delta \phi} \delta \phi
\nonumber \\
& & \nonumber \\
& &
+\frac{\delta S}{ \delta {\Omega}_{(-)}{}^{{a}}{}_{{b}}}
\left(
\frac{\delta {\Omega}_{(-)}{}^{{a}}{}_{{b}}}{\delta g_{\mu\nu}}\delta g_{\mu\nu}
+\frac{\delta {\Omega}_{(-)}{}^{{a}}{}_{{b}}}{\delta B_{\mu\nu}} \delta B_{\mu\nu}
+\frac{\delta {\Omega}_{(-)}{}^{{a}}{}_{{b}}}{\delta A^{A_{i}}{}_{\mu}}\delta
A^{A_{i}}{}_{\mu}
\right)\, .
\end{eqnarray}

We can then apply a lemma proven in Ref.~\cite{Bergshoeff:1989de}: $\delta
S/\delta {\Omega}_{(-)}{}^{{a}}{}_{{b}}$ is proportional to $\alpha'$ and to
the zeroth-order equations of motion of $g_{\mu\nu},B_{\mu\nu}$ and $\phi$
plus terms of higher order in $\alpha'$.

The upshot is that, if we consider field configurations which solve the
zeroth-order equations of motion\footnote{These can be obtained from
  Eqs.~(\ref{eq:eq1})-(\ref{eq:eq4}) by setting $\alpha'=0$. This eliminates
  the Yang-Mills fields, the $T$-tensors and the Chern-Simons terms in $H$.}
up to terms of order $\alpha'$, the contributions to the equations of motion
associated to the implicit variations are at least of second order in
$\alpha'$ and we can safely ignore them here.

If we restrict ourselves to this kind of field configurations, the equations
of motion reduce to 

\begin{eqnarray}
\label{eq:eq1}
R_{\mu\nu} -2\nabla_{\mu}\partial_{\nu}\phi
+\tfrac{1}{4}{H}_{\mu\rho\sigma}{H}_{\nu}{}^{\rho\sigma}
-T^{(2)}{}_{\mu\nu}
& = & 
0\, ,
\\
& & \nonumber \\
\label{eq:eq2}
(\partial \phi)^{2} -\tfrac{1}{2}\nabla^{2}\phi
-\tfrac{1}{4\cdot 3!}{H}^{2}
+\tfrac{1}{8}T^{(0)}
& = &
0\, ,
\\
& & \nonumber \\
\label{eq:eq3}
d\left(e^{-2\phi}\star {H}\right)
& = &
0\, ,
\\
& & \nonumber \\
\label{eq:eq4}
\alpha' e^{2\phi}\mathfrak{D}_{(+)}\left(e^{-2\phi}\star {F}^{A_{i}}\right)
& = & 
0\, ,
\end{eqnarray}

\noindent
where $\mathfrak{D}_{(+)}$ stands for the exterior derivative covariant with
respect to each $\mathrm{SU}(2)$ subgroup and with respect to the torsionful
connection $\Omega_{(+)}$: suppressing the subindices $1,2$ that distinguish
the two subgroups, it takes the explicit form 

\begin{equation}
\label{eq:eq5}
e^{2\phi}d\left(e^{-2\phi}\star {F}^{A}\right)
+\epsilon^{ABC}{A}^{B}\wedge \star F^{C}
+\star {H}\wedge{F}^{A}
= 
0\, .   
\end{equation}

If the ansatz is given in terms of the 3-form field strength, we also need to
solve the Bianchi identity 

\begin{equation}
\label{eq:BianchiH}
d{H}  
-
\tfrac{1}{3}T^{(4)}
=
0\, ,
\end{equation}

\noindent
as well. 

\section{Solution of the equations for the corrections}
\label{sec-solutioncorrections}

In this Appendix we are going to show how we have solved the equations
of motion of the Heterotic Superstring effective field theory to first
order in $\alpha'$ using the ansatz Eqs.~(\ref{eq:ansatzcorrections1})
and (\ref{eq:ansatzcorrections2}), which describes corrections to the
zeroth-order solution in Eqs.~(\ref{eq:RNBH}) and
(\ref{eq:aefunctions}) codified in the functions $\delta_{X}$ with
$X=A,B,C\cdots$.

Because of this formulation of our anstaz, we can apply the lemma of
Ref.~\cite{Bergshoeff:1989de} and, therefore, we only need to solve
Eqs.~(\ref{eq:eq1})-(\ref{eq:eq3}), since we are not going to
introduce 10-dimensional Yang-Mills fields. In all computations we
will ignore all terms of second order in $\alpha'$ and higher. We will
denote by $k_{i}$ the integration constants and the Einstein equations
will be denoted by $\mathcal{E}_{ab}$. The components of the
Kalb-Ramond 3-form field strength, the spin connection, the torsionful
spin connection and their curvatures, which are necessary to write the
equations for our ansatz, can be found in
Appendix~\ref{sec-connections}.

It is convenient to start by studying the equation of motion of the
Kalb-Ramond field Eq.~(\ref{eq:eq3}). Substitution of the ansatz gives 

\begin{equation}
e^{-2\phi}=\frac{k_{1}}{D r^{2}}\, ,
\end{equation}

\noindent
and, expanding the function $D$ and comparing with the expansion of
$e^{\phi}$, both in Eqs.~(\ref{eq:ansatzcorrections2}), we find that

\begin{equation}
\label{eq:deltaDdeltaphi}
k_{1}= e^{-2\phi_{\infty}}/p\, ,
  \hspace{1cm}
  2\delta_{\phi} = \delta_{D}r^{2}/p\, .
\end{equation}

Next, we consider the Bianchi identity of the Kalb-Ramond 3-form field
strength Eq.~(\ref{eq:BianchiH}). Substituting the ansatz, we obtain a
relation between $\delta_{E}$ and $\delta_{C}$ and a relation between
$\delta_{F}$ and $\delta_{D}$

\begin{align}
\label{eq:bianchi1}
\delta_{E} 
  & =
    -\frac{p}{r^{2}}\delta_{C}
    +\frac{p}{2}\frac{1-a^{2}}{r^{4}}-\frac{p^{3}/8}{r^{6}}+\frac{k_{2}}{r^{2}}\, ,
  \\
  & \nonumber\\
  \label{eq:bianchi2}
  \delta_{F}
  & =
    -\frac{pa^{2}}{2r^{3}}-\frac{r^{2} a}{p}\delta_{G} +k_{3}\, .
\end{align}

The integration constant $k_{2}$ corrects the value of the electric
and magnetic charges. Therefore, we will simply set $k_{2}=0$. As a
general rule, we will adjust the integration constants so that there
are no $\alpha'$ corrections of the fields at infinity. Thus, we also
set $k_{3}=0$.

The dilaton equation (\ref{eq:eq2}) gives the following relation
between $\delta_{D}$ and $\delta_{E}$:

\begin{equation}
\label{eq:dila}
\left[r^{2} a^{2} (r^{2} \delta_{D})'  \right]'
=
2 p^{2} (\delta_{D}-\delta_{E})+\frac{p}{16 r^{6}}
\left( 25 p^{4} -96 M p^{2} r+96 M^{2} r^{2}\right)\, ,
\end{equation}

\noindent
and from the Einstein equations we get the following relations:

\begin{itemize}
\item $\mathcal{E}_{04}$

\begin{equation}
 \label{eq:e04}
 \delta_{G}
 =
 \frac{a}{8} \left[ \frac4r (a^{2})'-\frac{p^{2}}{r^{4}}-\frac{r^{2}}{p}\delta_{F}' \right]'\, ,
 \end{equation}

\item $\mathcal{E}_{44}$
  
 \begin{equation}
\label{eq:e44}
\left(r^{2} a^{2} \delta_{C}'\right)'
=
p(\delta_{D}-\delta_{E})+\frac{p^{4}}{8r^{6}}\, ,
\end{equation}

\item $\mathcal{E}_{00}+\mathcal{E}_{11}$
\begin{equation}
\label{eq:e0011}
\left(\frac{\delta_{A}}{a}+a\delta_{B}\right)'
=
\frac{r}{2p}(p\delta_{C}-r^{2} \delta_{D})''-\frac{p^{2}}{4r^{5}}\, ,
\end{equation}

\item $\mathcal{E}_{22}$ and $\mathcal{E}_{33}$

\begin{equation}
\label{eq:22}
\frac{(r^{2} a^{5} \delta_{B})'}{a^{2}}
=
r^{2} a^{2} \left(\frac{\delta_{A}}{a}-r^{2}\frac{\delta_{D}}{p}+\delta_{C}\right)'
+pr\delta_{E}
-\frac{1}{4r^{5}}\left[ (2 p^{2}+12 M^{2}) r^{2}-14 M p^{2} r+3 p^{4}\right]\, ,
\end{equation}

\item $\mathcal{E}_{00}$
  
\begin{multline}
\label{eq:e00}
\frac{1}{a r^{2}}\left[r^{2} a^{3} \left( \frac{\delta_{A}}{a}
  \right)' \right]' =
\frac{(a^{2})'}{2p}(r^{2}\delta_{D}-p\delta_{C})'+\frac{p}{r^{2}}\delta_{D}
+\frac{1}{2(a^{2})' r^{4}}\left\{\left[r^{2} (a^{2})'\right]^{2} a \delta_{B}\right\}'\\
\\
+\frac{1}{8r^{8}}\left[6(4 M^{2}-p^{2})r^{2}-16 M p^{2} r+5
  p^{4}\right]\, .
 \end{multline}

\end{itemize}

These equations can be easily decoupled.  Substituting
Eq.~(\ref{eq:bianchi2}) in Eq.~(\ref{eq:e04}) gives a second order
equation for $\delta_{G}$.  Using the standard definition of $r_{+}$
and $r_{-}$ Eq.~(\ref{eq:rpm}) with $0<r_{-}<r_{+}<r$, imposing
reality and regularity on $r_{+}$ (it is not possible to have
regularity both on $r_{+}$ and $r_{-}$) and the above condition on the
vanishing of the corrections to the fields at infinity, we find

\begin{multline}
  \delta_{G}
  =
  \frac{a}{24r^{5} r_{-}^{2} r_{+}^{3} (r-r_{-})}
  \left\{r_{-} \left[-6 r^{3} \left(r_{-}^{4}+3 r_{-}^{2} r_{+}^{2}
        +2 r_{-} r_{+}^{3}+2 r_{+}^{4}\right)
 +6 r^{2} r_{-} r_{+}^{2} \left(3  r_{-}^{2}+r_{+}^{2}\right)\right.\right.\\
 \left.+4 r r_{-}^{2} r_{+}^{3} (3 r_{-}-7   r_{+})+40 r_{-}^{3} r_{+}^{4}\right]\\
   \left.-12 r^{3} r_{+} (r-r_{-}) (r_{-}+r_{+})
   \left(r_{-}^{2}+r_{+}^{2}\right)  \log \left(1-\frac{r_{-}}{r}\right)\right\}
   +\frac{k_{(4)}}{r^{2}}\, .
\end{multline}

\noindent
Then, using this result in Eq.~(\ref{eq:bianchi2}) with $k_{3}=0$ we
get $\delta_{F}$ and we see that we must also set $k_{(4)}=0$.

Combining Eq.~(\ref{eq:dila}) and Eq.~(\ref{eq:e44}) gives

\begin{equation}
 \left[r^{2} a^{2} (r^{2} \delta_{D}-2p\delta_{C})'  \right]'=\frac{p}{16 r^{6}}
  \left( 21 p^{4} -96 M p^{2} r+96 M^{2} r^{2}\right)\,,
\end{equation}

\noindent
which can be integrated, giving

\begin{multline}
  \label{eq:deltaDdeltaC}
  r^{2}\delta_{D}-2p\delta_{C}
  =
   +\frac{1}{40 \sqrt{2} r^{4}
    (r_{-} r_{+})^{5/2}} \left\{4 r^{4} \left(r_{-}^{4}-9 r_{-}^{3}
      r_{+}+r_{-}^{2} r_{+}^{2}-9
      r_{-} r_{+}^{3}+r_{+}^{4}\right) \log\left(1-\frac{r_{-}}{r}\right)\right.\\
  \left.+\frac{r_{-} r_{+}}{4} \left[2 r^{3} (r_{-}+r_{+}) \left(r_{-}^{2}-10
        r_{-} r_{+}+r_{+}^{2}\right)+r^{2} r_{-} r_{+}
      \left(r_{-}^{2}-19 r_{-}
        r_{+}+r_{+}^{2}\right)\right.\right.\\
  \left.\left.-6 r r_{-}^{2} r_{+}^{2} (r_{-}+r_{+})+21 r_{-}^{3}
      r_{+}^{3}\right]\right\}\, .
\end{multline}

Using this relation to express $\delta_{D}$ in terms of $\delta_{C}$
in Eq.~(\ref{eq:e44}), and using Eq.~(\ref{eq:bianchi1}) to express
$\delta_{E}$ in terms of $\delta_{C}$ in Eq.~(\ref{eq:e44}), we obtain
a second order equation for $\delta_{C}$ solved by

\begin{multline}
  \label{eq:deltaC}
  \delta_{C}
  =
  \frac{1}{140 r^{2} r_{-}^{3} r_{+}^{3}}\left[r^{2} \left(9 r_{-}^{4}+74
   r_{-}^{3} r_{+}+51 r_{-}^{2} r_{+}^{2}+74 r_{-} r_{+}^{3}+9
   r_{+}^{4}\right)\right.\\
   \left.+2 r_{-} r_{+}
     \left(34 r_{-}^{2}+23 r_{-} r_{+}+34 r_{+}^{2}\right)
     \left( r_{-} r_{+}-r (r_{-}+r_{+}) \right)\right] \log
   \left(1-\frac{r_{-}}{r}\right)\\
   +\frac{1}{560 r^{4}}\left\{\frac{580 r^{3} r_{-}}{r_{+}^{2}}+\frac{12
   r^{2} \left(595 r r_{-}+143 r r_{+}+226 r_{-}^{2}+83 r_{-}
   r_{+}\right)}{r_{-}^{2}+4 r_{-} r_{+}+r_{+}^{2}}-\frac{2 r^{2} (756
   r+535 r_{-})}{r_{+}}\right.\\
   \left.+4 r \left(\frac{74 r^{2}}{r_{-}}-77 r+37
   r_{-}\right)+r_{+} \left(\frac{36 r^{3}}{r_{-}^{2}}-\frac{254
   r^{2}}{r_{-}}+148 r+101 r_{-}\right)\right\}\, .
\end{multline}

Using this result in Eq.~(\ref{eq:bianchi1}) with $k_{2}=0$ we get
$\delta_{E}$ and using it in Eq.~(\ref{eq:deltaDdeltaC}) we get
$\delta_{D}$. The latter gives us $\delta_{\phi}$ via
Eq.~(\ref{eq:deltaDdeltaphi}).

Only $\delta_{A}$ and $\delta_{B}$ remain to be determined. We could
integrate Eq.~(\ref{eq:e0011}) to get $\delta_{A}$ in terms of
$\delta_{B}$, and, substituting everything in Eq.~(\ref{eq:22}), we
could obtain a first order equation for $\delta_{B}$ which could also
be immediately integrated.
However, given that the dilaton and Kaluza-Klein scalars become
non-trivial when the $\alpha'$ corrections are taken into account and
given that the Einstein metric includes certain powers of them, it is
more convenient to use variables different from $A$ and $B$ to
describe the metric. As a matter of fact, some of the equations take a
much simpler form in terms of those variables.

We define two new variables $N$ and $f$ and a new radial coordinate
$\rho$ from the 4-dimensional Einstein metric, given by

\begin{equation}
  ds^{2}_{(4)}
  =
  C e^{-2(\phi-\phi_{\infty})}\left[A^{2} dt^{2}-B^{2} dr^{2}-r^{2} d\Omega^{2}_{(2)}\right]
\equiv
N^{2} f dt^{2}-\frac{d\rho^{2}}{f}-\rho^{2} d\Omega^{2}_{(2)}\, ,
\end{equation}

\noindent
and we define the $\alpha'$ corrections to $N$ and $f$ by 

\begin{equation}
  N^{2}=1+\alpha' \delta_{N}\, ,
  \qquad
  f=a^{2}(r)+\alpha' \tilde\delta_{f}=a^{2}(\rho)+\alpha'\delta_{f}\, .
\end{equation}

\noindent
The corrections $\delta_{N}$, $\delta_{f}$ and $\tilde{\delta}_{f}$ are
related to the other corrections defined before by

\begin{align}
  \delta_{N}
  & =
    \left( \delta_{C}-r^{2}\frac{\delta_{D}}{p} \right)
 -r \left( \delta_{C}-r^{2}\frac{\delta_{D}}{p} \right)'
    +2\left( \frac{\delta_{A}}{a}+a\delta_{B} \right)\, ,
  \\
  \nonumber \\
  \tilde\delta_{f}
  & =
    a^{2}\left[r \left( \delta_{C}-r^{2}\frac{\delta_{D}}{p} \right)'-2 a\delta_{B}  \right]\, .
  \\
  \nonumber \\
  \delta_{f}
  & =
    \tilde\delta_{f}-\frac{r}{2}(a^{2})'
    \left( \delta_{C}-r^{2}\frac{\delta_{D}}{p} \right)\, .
\end{align}

As we have advanced, some of the above equations simplify when
expressed in terms of $\delta_{N}$ and $\delta_{f}$, namely:

\begin{itemize}
 \item $\mathcal{E}_{00}+\mathcal{E}_{11}$

\begin{equation}
  \delta_{N}'+ \frac{p^{2}}{2r^{5}}=0\,,
 \end{equation}

\item $\mathcal{E}_{00}+\mathcal{E}_{22}$

\begin{equation}
  \delta_{f}''-\frac{2}{r^{2}}\delta_{f}=
  -\frac{p}{4r^{8}}\left( 11p^{3}-42 M p r+18 p r^{2}
  \right)\,,
\end{equation}

\noindent
where we substituted the expression (\ref{eq:bianchi1}) for
$\delta_{E}$ woth $k_{2}=0$.

\end{itemize}

These equations can be easily integrated to give

\begin{eqnarray}
  \delta_{N}
  & = &
        \frac{p^{2}/8}{r^{4}}\, ,
  \\
  & & \nonumber \\
  \delta_{f}
  & = &
        -\frac{p^{2}/4}{r^{4}}
        \left(1 -\frac{3M/2}{r}+ \frac{11p^{2}/40}{r^{2}}
        \right) +r^{2} k_{5}
 +\frac{k_{6}}{r}\,.
\end{eqnarray}

\noindent
We can set $k_{6}=0$ because that integration constant simply
renormalizes the mass. As for $k_{5}$, substituting the expressions we
have found for the $\delta$s in Eq.~(\ref{eq:22}) (or equivalently
(\ref{eq:e00})), one finds that $k_{5}=0$.

Observe that, since the new radial coordinate
$\rho=r+\alpha'\delta_{\rho}$, $r$ can be replaced by $\rho$ in all
the $\alpha'$-correction functions $\delta_{X}$

\section{Connections and curvatures}
\label{sec-connections}

Our ansatz for the metric is 

\begin{equation}
ds^{2} 
= 
A^{2}dt^{2} - B^{2}dr^{2} - r^{2}[d\theta^{2}+\sin^{2}{\theta}d\phi^{2}]
-C^{2}[dz+Fdt]^{2} - - d\vec{y}^{\, 2}_{5}\, ,
\end{equation}

\noindent
where $A,B,C,F$ are functions of the coordinate $r$. The expansions of these
functions in powers of $\alpha'$ are assumed to be of the form

\begin{equation}
A\sim a+\alpha' \delta_{A}\, ,\,\,\,\,\,\,\,
B \sim a^{-1}+\alpha' \delta_{B}\, ,\,\,\,\,\,\,
C\sim 1 +\alpha' \delta_{C}\, ,\,\,\,\,\,\,\,
F\sim \alpha' \delta_{F}\, ,  
\end{equation}

\noindent
and, since we are only interested in keeping terms of zeroth and first orders
in $\alpha'$, at some point we will discard terms such as $C'F,F^{2}$ etc.

In the obvious Vielbein basis

\begin{equation}
\label{eq:Vielbein}
e^{0} = Adt\, ,\,\,\,\,
e^{1} = Bdr\, ,\,\,\,\,
e^{2} = rd\theta\, ,\,\,\,\,  
e^{3} = r\sin{\theta}d\phi\, ,\,\,\,\,  
e^{4} = C[dz+Fdt]\, ,\,\,\,\,
e^{i}= dy^{i}\, ,  
\end{equation}

\noindent
the only non-vanishing components of the spin connection ($de^{a}=
\omega^{a}{}_{b}\wedge e^{b}$) are

\begin{equation}
  \begin{array}{rclrclrcl}
\omega^{0}{}_{1} & = & -{\displaystyle\frac{A'}{AB}} e^{0}+{\displaystyle\frac{CF'}{2AB}} e^{4}\, ,\hspace{1cm} &  
\omega^{0}{}_{4} & = &  {\displaystyle\frac{CF'}{2AB}} e^{1}\, ,\hspace{1cm} &  
\omega^{1}{}_{2} & = &  {\displaystyle\frac{1}{Br}} e^{2}\, , 
\\
& & & & & & & & \\
\omega^{1}{}_{3} & = &  {\displaystyle\frac{1}{Br}} e^{3}\, , &
\omega^{1}{}_{4} & = &  {\displaystyle\frac{CF'}{2AB}e^{0}+ \frac{C'}{BC}}
e^{4}\, , 
\hspace{1cm} & 
\omega^{2}{}_{3} & = &  {\displaystyle\frac{\cot \theta}{r}} e^{3}\, ,
\end{array}
\end{equation}

\noindent
or

\begin{equation}
  \begin{array}{rclrcl}
\omega^{0}{}_{1} & = &
{\displaystyle\left(-\frac{A'}{B}+\frac{C^{2}FF'}{2AB}\right)dt+\frac{C^{2}F'}{2AB}dz}\,
,
\hspace{1cm} &  
\omega^{0}{}_{4} & = &  {\displaystyle\frac{CF'}{2A}}dr\, ,\\
& & & & & \\
\omega^{1}{}_{2} & = &  {\displaystyle\frac{1}{B}} d\theta\, , &
\omega^{1}{}_{3} & = &  {\displaystyle\frac{\sin{\theta}}{B}}d\phi \, , \\
& & & & & \\
\omega^{1}{}_{4} & = &  {\displaystyle\left(\frac{CF'}{2B}  +\frac{C'F}{B}\right)dt+ \frac{C'}{B}}dz\, ,  & 
\omega^{2}{}_{3} & = &  {\displaystyle\cos{\theta}} d\phi\, .
\end{array}
\end{equation}

\noindent
Taking into account the above expansions in $\alpha'$ and keeping only terms
of up to first order in $\alpha'$, we can already simplify some terms:

\begin{equation}
FF'\sim C' F \sim C'F'\sim 0 +\mathcal{O}(\alpha'^{2})\, ,
\hspace{1.5cm}
CF'\sim (CF)' \sim F' +\mathcal{O}(\alpha'^{2})\, ,
\end{equation}

\noindent
and, to this order, we can replace the above components of the spin connection
1-form by 

\begin{equation}
  \begin{array}{rclrcl}
\omega^{0}{}_{1} & = & {\displaystyle -\frac{A'}{B}dt+\frac{F'}{2AB}dz }\, ,
\hspace{1cm} &  
\omega^{0}{}_{4} & = &  {\displaystyle\frac{F'}{2A}}dr\, ,\\
& & & & & \\
\omega^{1}{}_{2} & = &  {\displaystyle\frac{1}{B}} d\theta\, , &
\omega^{1}{}_{3} & = &  {\displaystyle\frac{\sin{\theta}}{B}}d\phi \, , \\
& & & & & \\
\omega^{1}{}_{4} & = &  {\displaystyle\frac{F'}{2B}dt+ \frac{C'}{B}}dz\, ,  & 
\omega^{2}{}_{3} & = &  {\displaystyle\cos{\theta}} d\phi\, .
\end{array}
\end{equation}

Using these components, the non-vanishing components of the curvature 2-form
($R^{a}{}_{a}=d\omega^{a}{}_{b}-\omega^{a}{}_{c}\wedge \omega^{c}{}_{b}$) can
be readily calculated:

\begin{equation}
\begin{array}{rcl}
R_{01}
& = &
{\displaystyle
\left(
\frac{A'}{B}
\right)'\, dt\wedge dr  
+
\left(\frac{F'}{2AB}\right)'
dr\wedge dz 
}
\\
& & \\  
& = &
{\displaystyle
\left(
\frac{A'}{B}
\right)'\frac{1}{AB}\, e^{0}\wedge e^{1}
+
\left(\frac{F'}{2AB}\right)'\frac{1}{B}\, 
e^{1}\wedge e^{4}\, , 
}
\\
& & \\
R_{02}
& = & 
{\displaystyle
\frac{A'}{B^{2}}dt\wedge d\theta +\frac{F'}{2AB^{2}}d\theta\wedge dz
}  
\\
& & \\  
& = &
{\displaystyle
\frac{A'}{AB^{2}r}\, e^{0}\wedge e^{2}+\frac{F'}{2AB^{2}r}\, e^{2}\wedge
e^{4}\, ,
}
\\
& & \\
R_{03}
& = & 
{\displaystyle
\frac{A'\sin{\theta}}{B^{2}}dt\wedge d\phi
+\frac{F'\sin{\theta}}{2AB^{2}}d\phi\wedge dz
}
\\
& & \\  
& = &
{\displaystyle
\frac{A'}{AB^{2}r}\, e^{0}\wedge e^{3}
+\frac{F'}{2AB^{2} r}\, e^{3}\wedge e^{4}\, ,
}
\\
& & \\
R_{04}
& = & 
{\displaystyle
\frac{A'C'}{B^{2}}dt\wedge dz 
= 
\frac{A'C'}{A B^{2}}e^{0}\wedge e^{4}\, , 
}
\\
& & \\
R_{12}
& = & 
{\displaystyle
\frac{B'}{B^{2}}dr\wedge d\theta
=
\frac{B'}{B^{3}r}\, e^{1}\wedge e^{2}\, 
}
\\
& & \\
R_{13}
& = & 
{\displaystyle
\frac{B'\sin{\theta}}{B^{2}}dr\wedge d\phi
=
\frac{B'}{B^{3}r}\, e^{1}\wedge e^{3}\, ,
}
\\
& & \\
R_{14}
& = & 
{\displaystyle
-\left( \frac{C'}{B} \right)' dr\wedge dz+\left( \frac{F'}{2 A B} \right)'A\, dt\wedge dr
}
\\
& & \\
& = &
{\displaystyle
-\left( \frac{C'}{B} \right)'\frac{1}{B}\, e^{1}\wedge e^{4}+\left( \frac{F'}{2 A B} \right)'\frac{1}{B}\, e^{0}\wedge e^{1}\, ,
}
\\
& & \\
R_{23}
& = & 
{\displaystyle
\frac{B^{2}-1}{B^{2}}\sin{\theta}d\theta\wedge d\phi
=
\frac{B^{2}-1}{B^{2}r^{2}}\, e^{2}\wedge e^{3}\, ,
}
\\
& & \\
R_{24}
& = & 
{\displaystyle
\frac{F'}{2B^{2}}dt\wedge d\theta
-
\frac{C'}{B^{2}}d\theta\wedge dz
}
=
{\displaystyle
\frac{F'}{2AB^{2}r}\, e^{0}\wedge e^{2}
-
\frac{C'}{B^{2}r}\, e^{2}\wedge e^{4}\, ,
}
\\
& & \\
R_{34}
& = & 
{\displaystyle
\frac{F'\sin{\theta}}{2B^{2}}dt\wedge d\phi
-
\frac{C'\sin{\theta}}{B^{2}}d\phi\wedge dz
}
=
{\displaystyle
\frac{F'}{2AB^{2}r}\, e^{0}\wedge e^{3}
-
\frac{C'}{B^{2}r}\, e^{3}\wedge e^{4}\, .
}
\end{array}
\end{equation}

The (flat) non-vanishing components of the Ricci tensor are

\begin{equation}
\begin{array}{rcl}
R_{00} 
& = & 
{\displaystyle
-\frac{1}{ABCr^{2}}\left(\frac{A'Cr^{2}}{B}\right)'\, ,
}
\\
& & \\   
R_{04}
& = & 
{\displaystyle
\frac{1}{2Br^{2}}\left(\frac{F'r^{2}}{AB}\right)'\, ,
}
\\
& & \\  
R_{11}
& = & 
{\displaystyle
\frac{1}{AB}\left(\frac{A'}{B}\right)'+\frac{2}{Br}\left(\frac{1}{B}  \right)'+\frac{1}{B}\left(\frac{C'}{B}  \right)'\, ,
}
\\
& & \\  
R_{22}=R_{33}
& = & 
{\displaystyle
\frac{1}{ABCr^{2}}\left(\frac{ACr}{B}\right)'-\frac{1}{r^{2}}\, ,
}
\\
& & \\  
R_{44}
& = & 
{\displaystyle
\frac{1}{ABr^{2}}\left(\frac{AC'r^{2}}{B}\right)'\, ,
}
\end{array}
\end{equation}

\noindent
and their expansion in $\alpha'$ takes the form

\begin{equation}
\begin{array}{rcl}
  R_{00} 
  & = & 
        {\displaystyle
        -\frac{1}{2r^{2}}\left[(a^{2})'r^{2}\right]'
        +\frac{\alpha'}{2r^{2}}
        \left\{
        \left[(a^{2})'r^{2}\right]'\left(\frac{\delta_{A}}{a}+a\delta_{B}+\delta_{C}\right)
        -2\left[ar^{2}(\delta_{A}'+a'\delta_{C}-aa'\delta_{B})\right]'
        \right\}
        \, ,
        }
  \\
  & & \\   
  R_{04}
  & = & 
        {\displaystyle
        \frac{\alpha' a}{2r^{2}}\left(\delta_{F}'r^{2}\right)'\, ,
        }
  \\
  & & \\   
  R_{11}
  & = & 
        {\displaystyle
        \frac{1}{2r^{2}}\left[(a^{2})'r^{2}\right]'
        -\alpha' 
        \left\{
        \tfrac{1}{2}(a^{2})'' \left(\frac{\delta_{A}}{a}+a\delta_{B}\right)
        -\left[a(\delta_{A}'-a'a\delta_{B}) \right]'
        \right.
        }
  \\
  & & \\
  & & 
        {\displaystyle
  \left.
  +\frac{2a^{2}}{r}(3a'\delta_{A}+a\delta_{B}')
  -a\left(a\delta_{C}'\right)'
  \right\}\, ,
  }
  \\
  & & \\
  R_{22}
  & = & 
        R_{33}
        =
        {\displaystyle
        \frac{1}{r^{2}}\left(a^{2} r\right)' -\frac{1}{r^{2}}
        +\alpha'
        \left\{
        -\frac{2a}{r^{2}}\left(a^{2} r\right)'\delta_{B} 
        +\frac{a^{2}}{r}\left(\frac{\delta_{A}}{a}-a\delta_{B}+\delta_{C} \right)' 
        \right\}\, ,
        }
  \\
  & & \\
  R_{44}
  & = &
        {\displaystyle
        \frac{\alpha'}{r^{2}}\left(a^{2}r^{2}\delta_{C}' \right)'\, .
        } 
\end{array}
\end{equation}

It is trivial to see that $a^{2}=1+k/r$ (the Schwarzschild solution) satisfies
the Einstein equations in vacuum $R_{ab}=0$ at zeroth order in $\alpha'$.

Our ansatz for the Kalb-Ramond 3-form field strength $H$ is

\begin{equation}
H
= 
D e^{0} \wedge e^{1}\wedge e^{4} 
+E e^{2} \wedge e^{3} \wedge e^{4}
+G e^{0} \wedge e^{2}\wedge e^{3}\, ,
\end{equation}

\noindent
so

\begin{equation}
H^{2} = 6(D^{2}+G^{2}-E^{2})\, .  
\end{equation}

The expansions of $D,E,G$ in powers of $\alpha'$ are assumed to be of the form

\begin{equation}
D\sim e +\alpha'\delta_{D}\, ,\,\,\,\,\,\,
E\sim e +\alpha'\delta_{E}\, ,\,\,\,\,\,\,
G\sim\alpha'\delta_{G}\, ,  
\end{equation}

\noindent
and, therefore, 

\begin{equation}
H^{2} = 12e\alpha' (\delta_{D}-\delta_{E})\, .  
\end{equation}

We only need to compute the $\Omega_{(-)\, ab}=
\omega_{ab}-\tfrac{1}{2}H_{cab}e^{c}$ connection to zeroth order in
$\alpha'$. The non-vanishing components are

\begin{equation}
\begin{array}{rclrclrcl}
\Omega_{(-)\, 01}
& = & 
-a' e^{0} -\tfrac{1}{2}e\, e^{4}\, ,\,\,\,\,
& 
\Omega_{(-)\, 04}
& = & 
\tfrac{1}{2}e\, e^{1}\, ,\,\,\,\,
&
\Omega_{(-)\, 12}
& = & 
{\displaystyle
-\frac{a}{r}\, e^{2}
}\, ,
\\
& & & & & & & & 
\\
\Omega_{(-)\, 13}
& = & 
{\displaystyle
-\frac{a}{r}\, e^{3}
}\, ,
&
\Omega_{(-)\, 14}
& = & 
{\displaystyle
-\tfrac{1}{2}e\, e^{0}
}\, ,
&
\Omega_{(-)\, 23}
& = & 
{\displaystyle
-\frac{\cot{\theta}}{r}\, e^{3}
-\tfrac{1}{2}e\, e^{4}
}\, ,
\\
& & & & & & & & 
\\
\Omega_{(-)\, 24}
& = & 
{\displaystyle
\tfrac{1}{2}e\, e^{3}
}\, ,
&
\Omega_{(-)\, 34}
& = & 
{\displaystyle
-\tfrac{1}{2}e\, e^{2}
}\, .
&
& & \\
\end{array}
\end{equation}

\noindent
and the non-vanishing components of its curvature 2-form are

\begin{equation}
\begin{array}{rcl}
R_{(-)}{}^{0}{}_{1} 
& \sim &
{\displaystyle 
\left[\tfrac{1}{2}(a^{2})'' -\frac{p^{2}}{4r^{4}} \right] e^{0} \wedge e^{1} +
\frac{pa}{r^{3}} e^{1} \wedge e^{4}\, , 
}
\\
& & \\
R_{(-)}{}^{0}{}_{2} 
& \sim & 
{\displaystyle 
- \frac{p^{2}}{4r^{4}} e^{1} \wedge e^{3} - \frac{pa}{2r^{3}} e^{2} \wedge
e^{4} + \frac{(a^{2})'}{2r} e^{0} \wedge e^{2}\, ,  
}
\\
& & \\
R_{(-)}{}^{0}{}_{3} 
& \sim &  
{\displaystyle 
\frac{p^{2}}{4r^{4}} e^{1} \wedge e^{2} - \frac{pa}{2r^{3}} e^{3} \wedge e^{4}
+ \frac{(a^{2})'}{2r} e^{0} \wedge e^{3}\, ,  
}
\\
& & \\
R_{(-)}{}^{0}{}_{4} 
& \sim &   
{\displaystyle 
-\frac{p^{2}}{4r^{4}} e^{0} \wedge e^{4}\, , 
}
\\
& & \\
R_{(-)}{}^{1}{}_{2} 
& \sim &   
{\displaystyle 
- \frac{p^{2}}{4r^{4}} e^{0} \wedge e^{3} + \frac{pa}{2r^{3}} e^{3} \wedge
e^{4} + \frac{(a^{2})'}{2r} e^{1} \wedge e^{2}\, ,  
}
\\
& & \\
R_{(-)}{}^{1}{}_{3} 
& \sim & 
{\displaystyle 
\frac{p^{2}}{4r^{4}} e^{0} \wedge e^{2} - \frac{pa}{2r^{3}} e^{2} \wedge e^{4}
+ \frac{(a^{2})'}{2r} e^{1} \wedge e^{3}\, ,  
}
\\
& & \\
R_{(-)}{}^{1}{}_{4} 
& \sim &   
{\displaystyle 
- \frac{p^{2}}{4r^{4}} e^{1} \wedge e^{4} + \frac{pa}{r^{3}} e^{0} \wedge
e^{1} + \frac{pa}{r^{3}} e^{2} \wedge e^{3}\, , 
}
\\ 
& & \\
R_{(-)}{}^{2}{}_{3} 
& \sim & 
{\displaystyle 
\left (\frac{a^{2}}{r^{2}}-\frac{1}{r^{2}} +  \frac{p^{2}}{4r^{4}} \right)
e^{2} \wedge e^{3} - \frac{pa}{r^{3}} e^{1} \wedge e^{4}\, , 
}
\\
& & \\
R_{(-)}{}^{2}{}_{4} 
& \sim &  
{\displaystyle 
\frac{p^{2}}{4r^{4}} e^{2} \wedge e^{4} - \frac{pa}{2r^{3}} e^{0} \wedge e^{2}
+ \frac{pa}{2r^{3}} e^{1} \wedge e^{3}\, , 
}
\\ 
& & \\
R_{(-)}{}^{3}{}_{4} 
& \sim & 
{\displaystyle 
\frac{p^{2}}{4r^{4}} e^{3} \wedge e^{4} - \frac{pa}{2r^{3}} e^{0} \wedge e^{3}
- \frac{pa}{2r^{3}} e^{1} \wedge e^{2}\, ,
}
\\ 
\end{array}
 \end{equation}

\noindent
or

\begin{equation}
\begin{array}{rclrclrcl}
R_{(-)\, 0101}
& = & 
\tfrac{1}{2}(a^{2})'' -{\displaystyle \frac{p^{2}}{4r^{4}} }\, ,\,\,\,\,\,
&
R_{(-)\, 0114}
& = & 
{\displaystyle
-\frac{pa}{r^{3}}\, ,
}
&
R_{(-)\, 0202}
& = & 
{\displaystyle
\frac{(a^{2})'}{2r}\, ,
}
\\
& & & & & & & & \\
R_{(-)\, 0213}
& = & 
-{\displaystyle \frac{p^{2}}{4r^{4}} }\, ,\,\,\,\,\,
&
R_{(-)\, 0224}
& = & 
{\displaystyle
\frac{pa}{2r^{3}}\, ,
}
&
R_{(-)\, 0303}
& = & 
{\displaystyle
\frac{(a^{2})'}{2r}\, ,
}
\\
& & & & & & & & \\
R_{(-)\, 0312}
& = & 
{\displaystyle \frac{p^{2}}{4r^{4}} }\, ,
&
R_{(-)\, 0334}
& = & 
{\displaystyle
\frac{pa}{2r^{3}}\, ,
}
&
R_{(-)\, 0404}
& = & 
-{\displaystyle \frac{p^{2}}{4r^{4}} }\, ,
\\
& & & & & & & & \\
R_{(-)\, 1203}
& = & 
{\displaystyle \frac{p^{2}}{4r^{4}} }\, ,
&
R_{(-)\, 1212}
& = & 
{\displaystyle
-\frac{(a^{2})'}{2r}\, ,
}
&
R_{(-)\, 1234}
& = & 
{\displaystyle
\frac{pa}{2r^{3}}\, ,
}
\\
& & & & & & & & \\
R_{(-)\, 1302}
& = & 
-{\displaystyle \frac{p^{2}}{4r^{4}} }\, ,
&
R_{(-)\, 1313}
& = & 
{\displaystyle
-\frac{(a^{2})'}{2r}\, ,
}
&
R_{(-)\, 1324}
& = & 
{\displaystyle
-\frac{pa}{2r^{3}}\, ,
}
\\
& & & & & & & & \\
R_{(-)\, 1401}
& = & 
{\displaystyle
\frac{pa}{r^{3}}\, ,
}
&
R_{(-)\, 1414}
& = & 
{\displaystyle \frac{p^{2}}{4r^{4}} }\, ,
&
R_{(-)\, 1423}
& = & 
{\displaystyle
\frac{pa}{r^{3}}\, ,
}
\\
& & & & & & & & \\
R_{(-)\, 2314}
& = & 
{\displaystyle
-\frac{pa}{r^{3}}\, ,
}
&
R_{(-)\, 2323}
& = & 
{\displaystyle
\frac{1-a^{2}}{r^{2}} -{\displaystyle \frac{p^{2}}{4r^{4}} }\, ,
}
&
R_{(-)\, 2402}
& = & 
{\displaystyle
-\frac{pa}{2r^{3}}\, ,
}
\\
& & & & & & & & \\
R_{(-)\, 2413}
& = & 
{\displaystyle
\frac{pa}{2r^{3}}\, ,
}
&
R_{(-)\, 2424}
& = & 
-{\displaystyle \frac{p^{2}}{4r^{4}} }\, ,
&
R_{(-)\, 3403}
& = & 
{\displaystyle
-\frac{pa}{2r^{3}}\, ,
}
\\
& & & & & & & & \\
R_{(-)\, 3412}
& = & 
{\displaystyle
-\frac{pa}{2r^{3}}\, ,
}
&
R_{(-)\, 3434}
& = & 
-{\displaystyle \frac{p^{2}}{4r^{4}} }\, .
&
& &  \\
\end{array}
\end{equation}

Then, for $a^{2}$ as given in Eq.~(\ref{eq:aefunctions})

\begin{eqnarray}
R_{(-)}{}_{abcd} R_{(-)}{}^{abcd}  
& \sim & 
4\left[  \left(- \frac{p^{2}}{4r^{4}}+ \frac12 (A^{2})''  \right)^{2} + \left(
    \frac{(A^{2})'}{r} \right)^{2} +  \left( \frac{p^{2}}{4r^{4}}
    -\frac{1}{r^{2}}+ \frac{A^{2}}{r^{2}}  \right)^{2}\right] 
\nonumber \\
& & \nonumber \\
& = & 
\frac{25p^{4}/2}{r^{8}} - \frac{48 p^{2} M}{r^{7}} + 
\frac{48M^{2}}{r^{6}}\, .
\end{eqnarray}


\end{document}